%% file: PatternListener.tex
\PassOptionsToPackage{unicode}{hyperref}
\PassOptionsToPackage{naturalnames}{hyperref}
\documentclass[sigconf]{acmart}
\usepackage{zi4}
\usepackage{subfigure}
\usepackage[ruled]{algorithm2e}
\usepackage{verbatim}
\usepackage{pifont}
\usepackage{balance}
\usepackage{url}
\usepackage{color}
\usepackage{makecell,epstopdf}
\usepackage{multirow}
\usepackage{fancyhdr}
\usepackage{float}

\newcommand{\mysize}{\fontsize{8}{8}\selectfont}

\DeclareFontFamily{U}{mathx}{\hyphenchar\font45}
\DeclareFontShape{U}{mathx}{m}{n}{
      <5> <6> <7> <8> <9> <10>
      <10.95> <12> <14.4> <17.28> <20.74> <24.88>
      mathx10
      }{}
\DeclareSymbolFont{mathx}{U}{mathx}{m}{n}
\DeclareFontSubstitution{U}{mathx}{m}{n}
\DeclareMathAccent{\widecheck}{0}{mathx}{"71}
\DeclareMathAccent{\wideparen}{0}{mathx}{"75}

\setcopyright{none}

\def\redl#1{{\color{black}#1}}
\def\red2#1{{\color{black}#1}}

\begin{document}
\title{PatternListener: Cracking Android Pattern Lock Using \\Acoustic Signals}






\author{Man Zhou$^{1}$, Qian Wang$^{1}$, Jingxiao Yang$^{1}$, Qi Li$^{2}$, Feng Xiao$^{1}$, Zhibo Wang$^{1}$, Xiaofeng Chen$^{3}$}
\affiliation{$^{1}$School of Cyber Science and Engineering, Wuhan University, China\\
             $^{2}$Graduate School at Shenzhen \& Department of Computer Science, Tsinghua University, China\\
             $^{3}$School of Cyber Engineering, Xidian University, China
             \\
\{zhouman,qianwang,yangjingxiao,f3i,zbwang\}@whu.edu.cn, qi.li@sz.tsinghua.edu.cn, xfchen@xidian.edu.cn
}

\begin{abstract}
Pattern lock has been widely used for authentication to protect user privacy on mobile devices (e.g., smartphones and tablets). 
Several attacks have been constructed to crack the lock. However, these approaches require the attackers to either be physically close to the target device or be able to manipulate the network facilities (e.g., WiFi hotspots) used by the victims. Therefore, the effectiveness of the attacks is significantly impacted by the environment of mobile devices. Also, these attacks are not scalable since they cannot easily infer \red2{unlock} patterns of a large number of \red2{devices}.

Motivated by an observation that fingertip motions on the screen of a mobile device can be captured by analyzing surrounding acoustic signals on it, we propose PatternListener\footnote{\bf This paper was accepted by the 25th ACM Conference on Computer and Communications Security (CCS). A preliminary version was submitted to the 24th ACM Conference on Computer and Communications Security (CCS) on May 19, 2017 and the 27th USENIX Security Symposium on Feb 8, 2018.}, a novel acoustic attack that cracks pattern lock by analyzing imperceptible acoustic signals reflected by the fingertip. It leverages speakers and microphones of the victim's device to play imperceptible audio and record the acoustic signals reflected \red2{by} the fingertip. In particular, it infers each unlock pattern by analyzing individual lines that \red2{compose} the pattern and are the trajectories of the fingertip.  
We propose several algorithms to construct signal segments \red2{according to the captured signals} for each line and infer possible candidates of each individual line according to the signal segments. 
Finally, we map all line candidates into grid patterns and thereby obtain the candidates of the entire unlock pattern. We implement a PatternListener prototype by using off-the-shelf smartphones and thoroughly evaluate it using 130 unique patterns. The real experimental results demonstrate that PatternListener can successfully exploit over 90\% patterns \red2{within} five attempts.
\end{abstract}

%
%



\maketitle



\textbf{\mysize ACM Reference Format:}\\
{\mysize Man Zhou, Qian Wang, Jingxiao Yang, Qi Li, Feng Xiao, Zhibo Wang, Xiaofeng Chen. 2018. PatternListener: Cracking Android Pattern Lock Using Acoustic Signals. In \textit{2018 ACM SIGSAC Conference on Computer and Communications Security (CCS'18), October 15--19, 2018, Toronto, ON, Canada.}
ACM, NewYork, NY, USA, 13 pages. https://doi.org/10.1145/3243734.3243777}

\section{Introduction}

\red2{Graphical information of pattern lock is particularly suitable for human brain, while mobile users always consider the limited digit PIN code unsafe~\cite{de2005picture}. Therefore,} pattern lock has been widely used to authenticate users on mobile devices. Users need to draw a pattern on the devices within seconds before using the devices, which enables an easy mechanism for user authentication.
According to a recent survey~\cite{ye2017cracking}, around 40\% of the participants use pattern lock as the screen lock to protect their devices, while 33\% of those who do not use it to lock their mobile systems often use pattern lock for identity authentication on apps, e.g., Alipay~\cite{Aliplay}.

Pattern lock security has attracted intensive attention recently. Many security mechanisms have been developed to ensure that the screen of mobile devices cannot be captured by other applications when users draw patterns. For example, sandbox and TrustZone provide software and hardware isolation for sensitive information (e.g., unlock pattern). All applications (app based or web based) will be constrained by such mechanisms so that they cannot access other private resources that \red2{are not assigned} to them. Hence, these mechanisms make 
\redl{the traditional attacks, e.g., hijacking the unlock screen or constructing phishing attacks, difficult to infer pattern lock.}
However, it is worth noting that applications on mobile devices can still access certain shared hardware resources like the accelerometer, camera, microphone, and GPS. Such resources may open a door to \red2{infer unlock pattern by using side channel information generated by them.} 

A large number of attacks~\cite{cai2011touchlogger,miluzzo2012tapprints,owusu2012accessory,simon2013pin,li2016csi} have been widely developed to crack PINs \red2{by capturing the features during typing PIN number.} \red2{For instance,} malware installed on a victim's device can identify the location of screen taps by leveraging the motion sensors (i.e., accelerometer and gyroscope) when users type on the soft keyboard on their devices~\cite{cai2011touchlogger,miluzzo2012tapprints,owusu2012accessory}. 
These approaches usually regard an entire unlocking process with multiple taps as a single sample for feature extraction and rely on abundant training with labeled data \red2{to perform} machine learning based analysis. In addition, the pressure of typing and changes of device orientation will significantly affect the attack accuracy. Simon et al.~\cite{simon2013pin} leveraged the microphone to detect touch events and the front camera to estimate the smartphone's orientation changes, and then correlated the changes to the position of the digit tapped by the victim. However, \red2{the estimation of orientation changes is impacted  by} ambient lighting and camera shake. Li et al.~\cite{li2016csi} considered that an attacker controls a public WiFi access point and inferred the keystrokes on the smartphone through WiFi CSI data, but WiFi signals are prone to be disrupted by nearby moving objects. 
\red2{These approaches cannot be applied to infer pattern lock. Actually, it is more difficult to crack pattern lock }since moving a fingertip on the screen during the pattern unlock process introduces much less disturbance to the mobile device than tapping numbers. Therefore, it is more challenging to infer the unlock pattern by leveraging the side channel information above.


Recently, several attacks \cite{aviv2010smudge,zhang2016privacy,ye2017cracking} have been constructed to crack pattern lock on Android. The Smudge attack~\cite{aviv2010smudge} leveraged oily residues left on the screen to infer the unlock pattern. However, the accuracy of inferring lock is highly impacted by the residues on the screen, which can be interfered by subsequent operations of users. Zhang et al.~\cite{zhang2016privacy} demonstrated the feasibility of inferring the pattern by analyzing wireless signals. Unfortunately, the proposed approach requires complicated network setup, and the effectiveness of the attack is easily \red2{interfered} by moving objects, e.g., \red2{the} people nearby. Ye et al.~\cite{ye2017cracking} cracked Android pattern lock by using video footage that records the victim's fingertip motions. It requires the attacker to be physically close enough to the device. Moreover, the attack accuracy is impacted by many physical factors, such as filming angle and distance, changes of light, and camera shake. \red2{In particular, these attacks cannot be used to infer unlock patterns of a large number of devices.}
In a nutshell, these existing attacks are not robust and scalable.


In this paper, we propose a novel acoustic attack, called PatternListener, to infer the sensitive unlock pattern by using imperceptible acoustic signals. The observation behind the attack is that the fingertip on the screen of a mobile device will reflect nearby acoustic signals, and the reflected signals embed the information of fingertip motions \red2{corresponding to the unlock pattern}. When a victim starts to draw his pattern, PatternListener
generates imperceptible audio and uses the speakers of the victim's device to play it, 
meanwhile, the microphones of the victim's device record the acoustic signals reflected by the fingertip. The recorded acoustic signals will be processed by a remote server to infer the fingertip patterns. PatternListerner constructs different lines according to the trajectories of the fingertip and infers each lock pattern by analyzing individual lines that compose the pattern.
Note that, 2D gesture tracking~\cite{nandakumar2016fingerio,wang2016device,yun2017strata} cannot be applied in PatternListener since they need simultaneously use two speaker-microphone pairs to track gestures, which requires re-configuring the smartphone systems and is not possible in our attack.

In particular, we utilize coherent detection together with static components removal to effectively eliminate noises in the signals, {and identify turning points of fingertip motions to accurately segment the acoustic signals into fragments associated with each line in the pattern.
We extract the movement features based on the changing trend of the path length of acoustic signals reflected by the moving fingertip so that we can infer the possible candidates of each line.}
We combine the candidates of different lines together to identify the most possible candidates for the unlock pattern. Note that, acoustic signals attenuate quickly as the distance increases, and thereby other irrelevant moving objects around, e.g., the victim's head, cannot interfere with the recorded acoustic signals, which means that PatternListener is robust to the interference from the environment. \red2{Particularly},  by collecting signals from various phone models, PatttenListener can easily infer unlock patterns of a large number of phone devices simultaneously.


The main contributions are summarized as follows:
\leftmargini=4mm
\begin{itemize}
\item We uncover a new vulnerability of pattern lock by leveraging speakers and microphones of mobile devices. To the best of our knowledge, this is the first work to  leverage speakers and microphones to reconstruct the victim's unlock pattern, which raises a serious issue that all shared hardware on phones can be leveraged to crack the security mechanisms.

\item We propose PatternListener, a novel attack to crack Android pattern lock by leveraging imperceptible acoustic signals. It is a more robust and  practical attack since it neither requires an attacker to be physically close to the victims nor is sensitive to the interference from the environment.

\item We develop several algorithms in PatternListener to infer lock patterns by analyzing acoustic signals reflected by the fingertip. Particularly, we recover each line constituting the pattern that is the trajectory of the fingertip drawing on the phone according to the signals. Therefore, PatternListener is scalable to analyze a large number of unlock patterns.
\item We implement a PatternListener prototype using off-the-shelf smartphones. The extensive experimental  results demonstrate that an attacker can successfully crack over 90\% of 130 patterns within five attempts. In particular, a complicated pattern with more lines cannot provide stronger protection for users under the attack of PatternListener. Moreover, PatternListener is robust to the changes of drawing speed and gestures, and different size of screens. It will not be significantly affected by surrounding objects and the ambient noise.
\end{itemize}


\input{attack-overview}
\input{system-design}

\input{evaluation}

{
\section{Countermeasures} \label{sec:defense}

\noindent {\bf Preventing Usage of Microphone in Background.} One straightforward countermeasure is to prevent the usage of microphone in the background during pattern drawing, and then the system can obstruct the access of microphone by any apps when a user is drawing pattern. However, it may incur a usability issue since many benign apps may still require using microphone even they are in the background.
{For example, a user may want to wake up and launch Google Assistant by saying ``Hey Google" or ``OK Google"~\cite{GoogleAssistant}.}
Note that, we do not notice any existing countermeasures that can effectively prevent our attack although the risk of abusing microphone has attracted more attention recently. In particular, in the newest Android version (i.e., Android 9.0~\cite{AndroidP}), the Android system prevents an app from using the microphones if the UID of the app is in an idle state. \redl{Unfortunately, it cannot effectively prevent the usage of microphone by apps in the background if they can be always active to use the microphone by running as an Android daemon, e.g., by using JobScheduler~\cite{JobScheduler}.}
Actually,
we evaluate the effectiveness of PatternListerner on a Pixel smartphone with Android 9.0 and find that PatternListerner can still effectively compromise the pattern lock.

\noindent {\bf Random Layout of Pattern Grids.} Another sophisticated defense is to randomize the layouts of the pattern grid. If the grid is shown in a different position with different space between the columns and rows each time during pattern drawing, the extracted movement features corresponding to the same pattern are also different. Thus, the attacker is not able to construct a valid ground-truth database and the attack will fail. However, similar to apps that enable random software keyboards, this countermeasure may impact the user experience since users are required to find each dot on a random layout of the pattern grid before pattern drawing.
}

\input{related-work}

\section{Conclusion} \label{sec:conclusion}
We presented PatternListener, a novel attack that reconstructs the unlock pattern by leveraging imperceptible acoustic signals. We implemented a PatternListener prototype using off-the-shelf smartphones. We evaluated PatternListener using the smartphones with 130 different patterns and the experimental results demonstrated that PatternListener achieved very high accuracy in reconstructing the unlock pattern on smartphones with various practical considerations. 
The experimental results showed that PatternListener is able to successfully crack over 90\% of the 130 patterns in five  attempts with only one sample for each pattern. Moreover, we can also draw several important conclusions from the experimental results: (1) {complicated pattern with more lines does not always mean stronger protection;} (2) the attack is more efficient if the device is held more stably; (3) PatternListener is relatively robust to the changes of drawing speed and different sizes of screens; (4) {surrounding objects and noise interference from} environment will not significantly affect the effectiveness of the attack.

\section*{Acknowledgments}
This work was supported in part by National Natural Science Foundation of China (NSFC) under Grant 61822207, U1636219, 61572278, U1736209, and 61572382, the Key Program of Natural Science Foundation of Hubei Province under Grant 2017CFA047 and 2017CFA007, and the China 111 Project under Grant B16037. Qian Wang and Qi Li are the corresponding authors of this paper.



\bibliographystyle{ACM-Reference-Format}
\bibliography{sigproc}

\section*{Appendix}

\subsection*{Permission of Accessing Microphone}\label{sec:Permission}

\begin{figure}[h]
\centering
\includegraphics[width=0.45\textwidth]{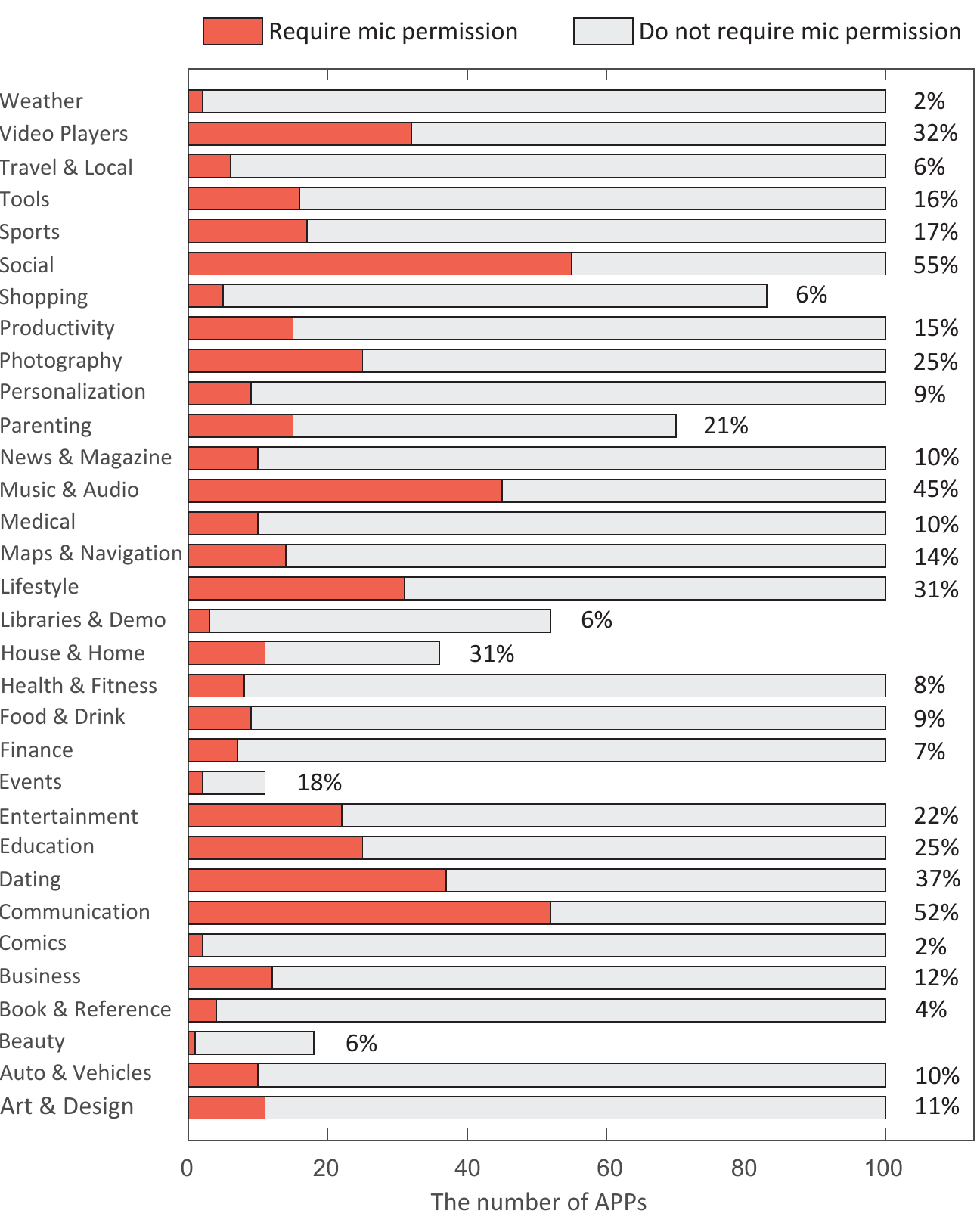}
\caption{The fraction of apps requiring the permission of accessing microphone in different categories.}
\label{fig:mic_permission}
\end{figure}

\begin{figure}[h]
\centering
\includegraphics[width=0.5\textwidth]{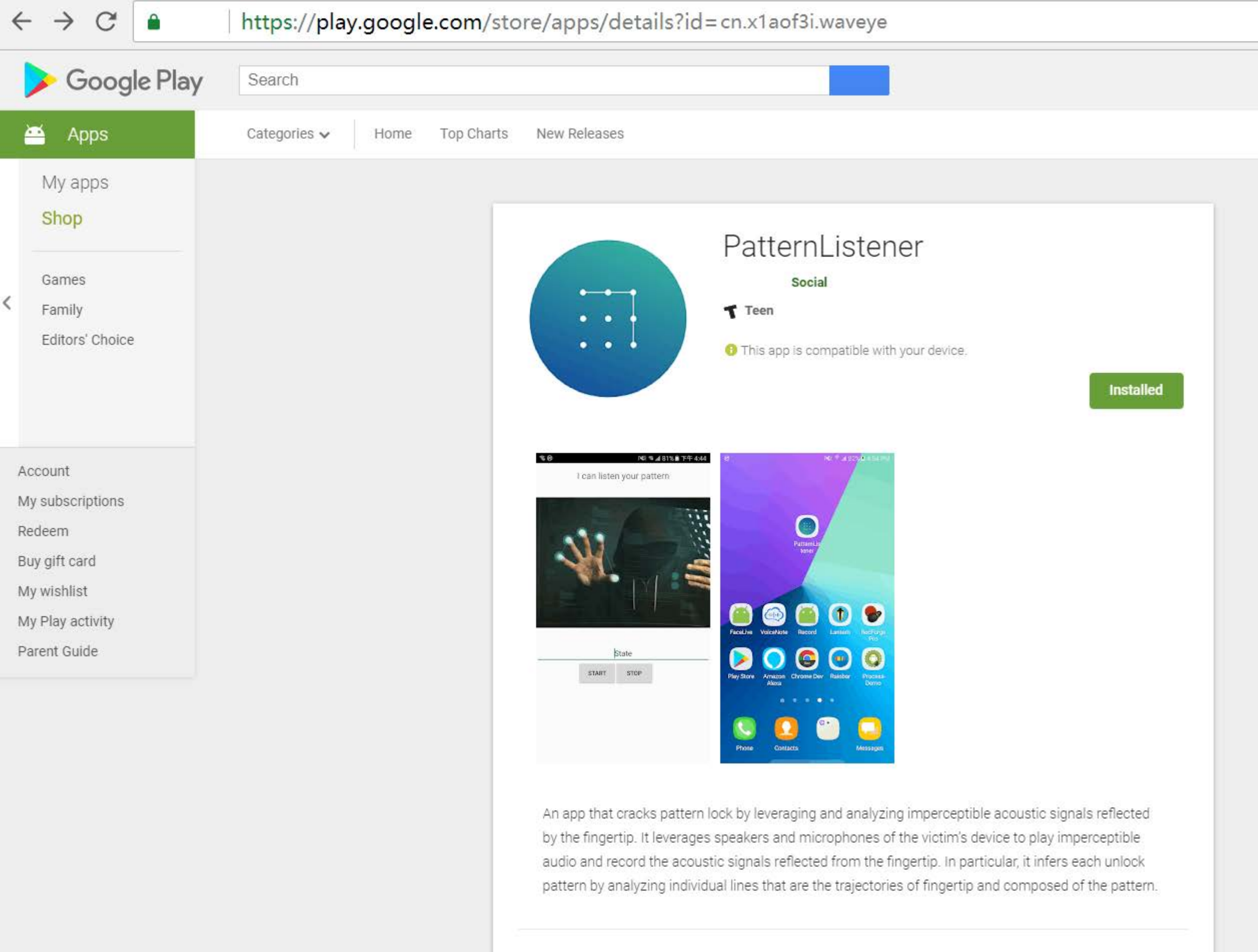}
\caption{The publishing page of PatternListener on Google Play.}
\label{fig:google}
\end{figure}

In order to successfully construct the attack, PatternListener requires the permissions to access the speaker, the microphone, and the motion sensors as well as the network access permission. Most permissions can be granted without user approval, except the permission of accessing microphone. We investigate the permission of accessing microphone in popular Android apps in the Google Play marketplace. We analyze the top 100 apps of each app category classified by Google Play. Note that, if the number of apps in a category is less than 100, we simply analyze all apps in the category. Figure \ref{fig:mic_permission} shows the fraction of apps requiring the permission of accessing microphone in different categories. We observe that the permission of accessing microphone is very popular in various Android apps. In particular, the permission of accessing microphone is required by 55\% social apps and 52\% communication apps. Thus, we can conclude that it is easy for PatternListener to obtain the permission upon installation after disguised as apps in these categories.
{
To evaluate the feasibility of PatternListener, we have submitted our PatternListener app to Google Play and assigned the social category to the app. The app passed the security check performed by Google and was published on Google Play. Figure \ref{fig:google} shows the screenshot of the Patternlistener app published on Google Play. To avoid being downloaded by users mistakenly, we have withdrawn the app from Google Play.
}

\end{document}

%% file: attack-overview.tex
\section{Cracking Pattern Lock}\label{sec:overview}
\subsection{Android Pattern lock}
\red2{Pattern lock is a typical lock policies to protect the sensitive information on users' devices. It authenticates users by asking them to draw a pattern on a given 3 $\times$ 3 grid, which is enabled in most mobile systems. 
Angeli et al.~\cite{de2005picture} report} that the human brain is particularly well-suited to remember such graphical information. There are increasing numbers of apps and OS providing pattern lock for their users as a protection option. In particular, pattern lock is widely applied in the Android ecosystem. A survey~\cite{aviv2014understanding} shows that among participants using Android devices, 257 of 354 (72\%) users used graphical passwords\red2{, and} 249 of 257 (97\%) users think pattern lock safe enough. Therefore, it is essential to study the security of the pattern lock mechanism.

\subsection{Threat Model}

In this paper, we study the vulnerability of pattern lock on Android by developing a novel attack called PatternListener. PatternListener aims to reconstruct unlock patterns of OS or apps on a victim's mobile device. It generates and plays imperceptible audio, meanwhile, the microphones of the victim's device record acoustic signals reflected by the fingertip, such that an attacker can analyze the recorded signals and reconstruct the pattern according to the fingertip motions. To ensure the feasibility of the attack, we will develop a malware that can be installed on mobile devices \red2{so} that an attacker can compromise many devices simultaneously and obtain the lock patterns. Note that, the malware runs in the background after being installed, which is similar to traditional malware~\cite{zhou2012dissecting}.  In order to launch the attack, PatternListener requires the permission to access speaker, microphone, and motion sensors (i.e., accelerometer and gyroscope) as well as network access permission.  Most permissions can be granted without user approval, except the permission of accessing the microphone. However, we observe that the permission of accessing microphone is very popular in Android apps. For instance,  microphone permissions are required by 55\% social apps and 52\% communication apps in the Google Play marketplace. \redl{The details can be found in the appendix.} Therefore, it is easy for PatternListener to obtain the permission after it is disguised as an app in these categories.

\redl{
Note that, PatttenListener \red2{can} crack pattern locks of various phone devices with different types of patterns simultaneously. The cracked patterns can be exploited in different ways. For example, PatttenListener can assign each cracked phone a unique serial number and then frequently broadcast the serial number through hidden acoustic signals~\cite{zhou2018dolphin}. The attacker can use a smartphone to detect and decode the hidden acoustic signals, and then understands which phones nearby have been cracked. Thus, the corresponding unlock pattern can be used to compromise the target device after the attacker has a chance to physically access the device for a short period of time.
}

\subsection{Overview of PatternListener}
\begin{figure}[!t]
\centering
\includegraphics[width=0.5\textwidth]{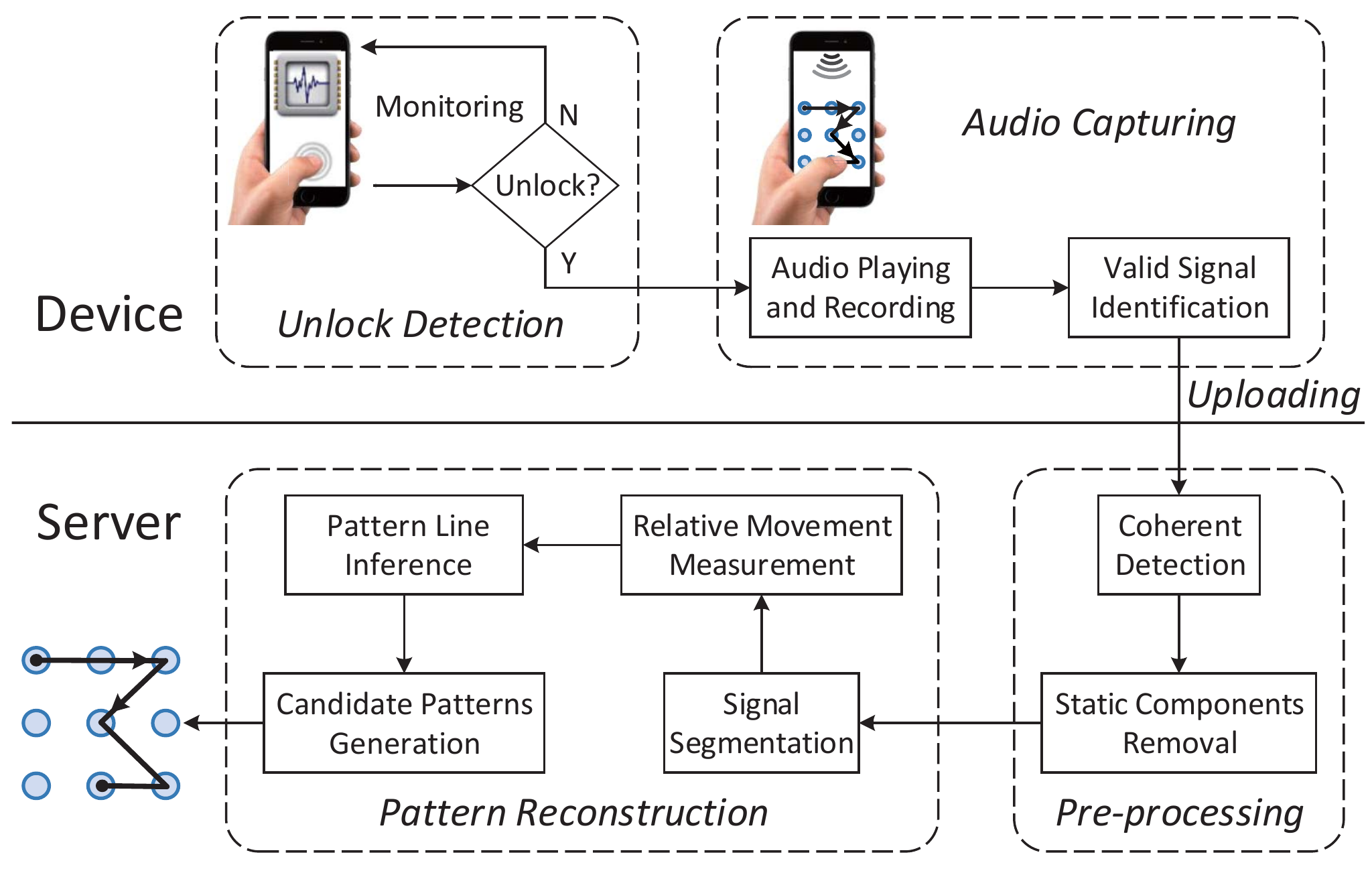}
\caption{Attack flow of PatternListener.}
\label{fig:attack_overview}
\vspace{-3mm}
\end{figure}

Figure \ref{fig:attack_overview} shows the flow of the attack constructed by PatternListener, which mainly consists of four phases: \textit{Unlock Detection, Audio Capturing, Pre-processing}, and \textit{Pattern Reconstruction}.

\noindent{\textbf{Unlock Detection}:}
This phase aims to detect when the victim is going to draw the unlock pattern. \red2{Thus} PatternListener can immediately play audio and record the reflected acoustic signals, and then captures the fingertip motions on the screen. In this paper, we consider \red2{two different unlock scenarios, i.e.,} screen unlock and app unlock. 

\noindent{\textbf{Audio Capturing}:}
This phase records acoustic signals to capture the fingertip motions on the screen during the unlock process. Once the unlock action is detected, PatternListener uses the speakers of the victim's device to play the generated imperceptible acoustic signals, and \red2{triggers} the microphones to record the acoustic signals reflected by the fingertip. The reflected acoustic signals corresponding to the unlock process will be \red2{identified} and uploaded to the server.

\noindent{\textbf{Pre-processing}:}
This phase extracts the sound signals corresponding to fingertip motions. In order to achieve this, PatternListener leverages the coherent detector to demodulate the baseband signals, and downsamples the signals to enable  efficient signal processing. \red2{Then, it} removes the static components to obtain the true acoustic signals reflected by the fingertip.

\noindent{\textbf{Pattern Reconstruction}:}
This phase finally reconstructs the victim's unlock pattern by analyzing the signals.  PatternListener analyzes the signals to obtain the trajectories of the fingertip drawing on the screen and recover the lines according to the trajectories. 
Since the pattern is composed of the lines, we can infer the candidate pattern by mapping the lines into grid patterns. \red2{It} includes four steps: the \textit{Signal Segmentation} step segments the acoustic signals into fragments, the {\em Relative Movement Measurement} step infers the movement of the fingertip,  the {\em Pattern Line Inference} step constructs lines representing the trajectories of the fingertip, and the \textit{Candidate Patterns Generation} step generates the candidate patterns according to the inferred lines.

%% file: system-design.tex
\section{PatternListener Design}\label{sec:design}

This section presents the detailed design of PatternListerner.

\subsection{Unlocking Detection}


Unlock detection aims to detect when the victim is going to draw the pattern so that PatternListener can immediately play audio and then record acoustic signals to capture the fingertip motions on the screen. We 
detect screen unlock and app unlock \red2{as follows.}

\noindent{\textbf{{Screen Unlock}:}
The screen of a device with pattern lock usually experiences the following three states when the victim is going to unlock the screen: (1) \textit{non-interactive}. the device is in sleep mode and the user \red2{cannot} interact with the device through the screen; (2) \textit{pre-interactive}. the screen is open and the user is waking up the device; (3) \textit{interactive}.  the device is totally activated, and the user can interact with the device through the screen. In \red2{the} Android system, the information of screen state will be automatically broadcasted when the state changes. Therefore, we can detect the action of screen unlock by monitoring \redl{the broadcasted information associated with} the state transition from \textit{non-interactive} to \textit{pre-interactive}.

\noindent{\textbf{App Unlock}:}
App unlock is different from screen unlock because it does not generate any broadcast information. In order to detect when \red2{a} victim is going to draw the app unlock pattern, we develop a simple \red2{and} effective scheme based on the following observation. The victim often have left or right swipes on the screen to find the app and click to select an app, and \red2{the fingertip motions} often pause for a few seconds \red2{before the unlock} because of the delay of app startup. 
These consecutive on-screen operations typically expose some spatial-temporal motion characteristics, which can be utilized to detect the action of app unlock. We leverage motion sensors to detect the click action on the screen. \red2{After the screen has been unlocked, we utilize the speaker to continuously play imperceptible audio. We can identify the swipe actions from the recorded acoustic signals since the moving fingertip will reflect the acoustic signals.} It is worth noting that the detection of swipe action is much easier than inferring the victim's unlock pattern by using acoustic signals because it does not need to accurately know the distance and angle of fingertip movements.

{\color{black}Note that, PatternListener does not need to capture the pattern unlock every time when the user draws the pattern. The pattern lock can be captured and cracked as long as we correctly detect the unlock action once. In fact, screen unlock action is very easy to be detected by the \red2{above} method because the victim tends to unlock the smartphone very frequently everyday. In this paper, for simplicity, we develop a simple unlock detection scheme, which is effective in detecting most app unlock behaviors. \red2{Actually,} 
we can possibly incorporate other suitable unlock detection mechanisms into PatternListener to improve the detection efficiency. For example, users may raise their smartphones before screen unlock (e.g., iOS 10 can detect raise action to wake up iPhone~\cite{RaisetoWake}) in most cases. Hence, we can leverage the motion sensor to detect the raise action of a smartphone and use \red2{the action} as a signal to detect that the victim is going to unlock the screen.}


\subsection{Audio Capturing}\label{sec:Capturing}

Once the unlock action is detected, PatternListener uses the speakers of the victim's device to play the generated imperceptible acoustic signals, and \red2{triggers} the microphones to record the acoustic signals reflected by the fingertip moving on the screen of a mobile device.
The reason why we leverage acoustic signals to reconstruct the unlock pattern is that the fingertip motions can be extracted by analyzing the reflected acoustic signals.

\begin{figure}[!t]
\centering
\includegraphics[width=0.7\columnwidth]{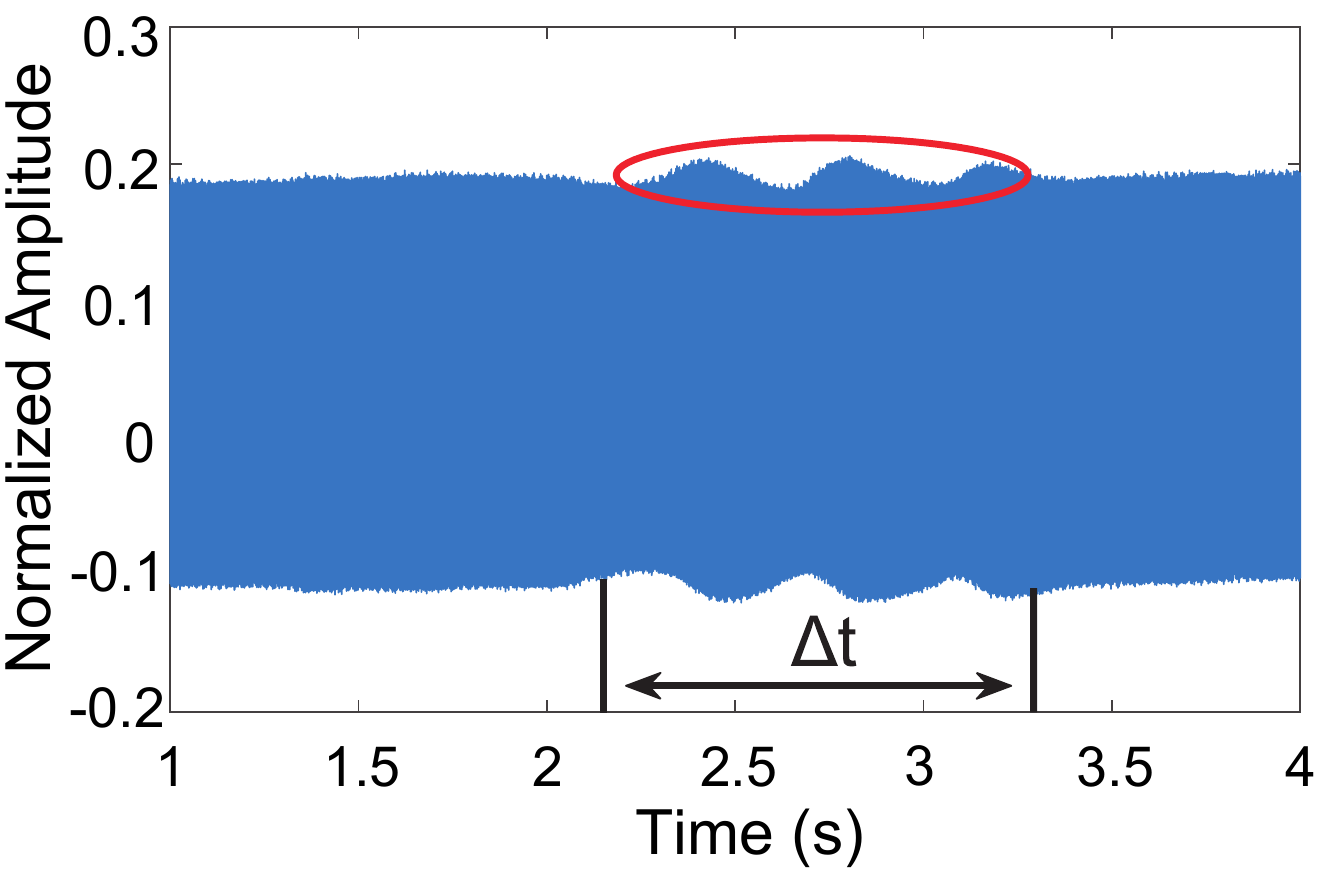}
\caption{An example of recorded acoustic signals. A fingertip moves on the screen of a smartphone during $\Delta t$.}
\label{fig:continuous-wave}
\vspace{-3mm}
\end{figure}

\vspace{1mm}
\noindent{\textbf{Audio Play with the Speaker}:}
The generated audio is a continuous wave acoustic signal of $A\sin 2\pi ft$, where $A$ is the amplitude and $f$ is the frequency of acoustic signals. The frequency $f$ is set to be in the range of $18 \sim 20$ kHz. The reason that we choose this frequency range is that the response frequency of most speakers and microphones is from $50$ Hz to $20$ kHz and most people cannot hear \red2{the} sound with a frequency higher than 18 kHz~\cite{wang2016messages}. {\color{black}Note that, some users may hear sound with a frequency higher than 18 kHz. However, we can lower the volume to make it almost imperceptible to them. Thereby, the generated audio can be recorded by the microphone but cannot be noticed by any users.} \red2{Moreover,} we observe that ambient noise becomes negligible at frequencies higher than $18$ kHz makes PatternListener undisturbed by ambient noise.

\vspace{1mm}
\noindent{\textbf{Audio Record with the Microphone}:}
The microphone \red2{records} the acoustic signals once the speaker plays the audio. The recorded acoustic signals capture the information of fingertip motions because the sliding fingertip on the screen of mobile devices reflects the played acoustic signals. Figure \ref{fig:continuous-wave} shows an example of recorded acoustic signals when the fingertip moves on the screen of a smartphone during $\Delta t$. We can observe that fingertip motions on the screen will lead to a significant interference to the acoustic signals (see the ellipse area shown in Figure \ref{fig:continuous-wave}).

\vspace{1mm}
\noindent{\textbf{\red2{Valid Signal Identification}}:} We are only interested in the acoustic signals corresponding to the unlock process. The unlock process starts when the fingertip touches the screen and terminates when the fingertip leaves the screen. The motion sensors can be used to detect the two key \red2{timepoints} and estimate the \red2{startpoint} of the pattern. The motion sensors data would change significantly when the fingertip clicks on the screen~\cite{miluzzo2012tapprints}. When the victim touches the screen, the finger gives a downward pressure to the phone, and the phone will rotate on the X-axis and Y-axis and move down on the Z-axis due to the pressure. When the fingertip leaves the screen, the pressure will disappear and the phone tends to return to its original location. That is, it will rotate on the X-axis and Y-axis and move up on the Z-axis. Such movements of the phone can be captured by motion sensors and thus  we can obtain the timepoints when the fingertip touches and leaves the screen by monitoring the changes of the data generated by the motion sensors. Then, the acoustic signals within the two key \red2{timepoints can be captured so that} an attacker can upload the signals to a server stealthily and analyze such signals to recover the unlock pattern.

\begin{figure}[!t]
\centering
\includegraphics[width=0.96\columnwidth]{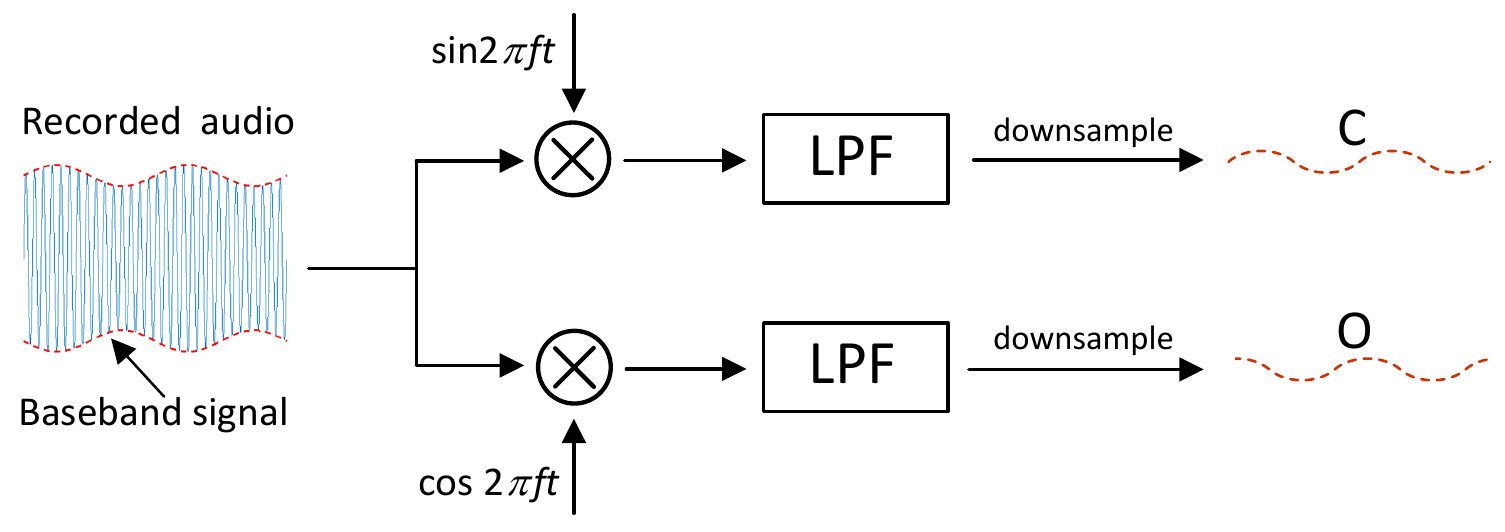}
\caption{The process of coherent detection.}
\label{fig:coherent-detector}
\vspace{-3mm}
\end{figure}

\subsection{Audio Preprocessing}\label{sec:Preprocess}
Before reconstructing the unlock pattern, PatternListener preprocesses the recorded audio to extract the sound signal component related to fingertip motions. PatternListener first leverages the traditional coherent {\color{black}detection~\cite{tse2005fundamentals}} to demodulate the baseband signals and downsamples the signals, and then removes the static components to obtain the true acoustic signals reflected by the fingertip.

\vspace{1mm}
\noindent{\textbf{Coherent Detection}:}  The played audio from the speaker can be treated as the carrier signal, and the signal related \red2{to} fingertip motions can be treated as the baseband signal. Thus, the recorded acoustic signals are the combination of the carrier signal and the baseband signal. The recorded acoustic signals are synchronized with the played acoustic signals, \red2{thus} we can \red2{utilize} the traditional coherent detector to demodulate the baseband signal from the recorded signals. The process of coherent detection is shown in Figure \ref{fig:coherent-detector}. Let $R(t)$ denote the recorded acoustic signals, $F_{lp}$ denote a low pass filter, and  $F_{ds}$ denote a downsampled function. Then the corresponding C (cophase) component and O (orthogonal) component are calculated as \red2{follows:}


\begin{equation}\label{eq:C/O}
\begin{aligned}
C(t) &= F_{ds}(F_{lp}(R(t)*A\sin 2\pi ft))\\
O(t) &= F_{ds}(F_{lp}(R(t)*A\cos 2\pi ft)).
\end{aligned}
\end{equation}

\begin{figure}[!t]
\centering
\subfigure[The unlock pattern ]{
\includegraphics[width=0.31\columnwidth]{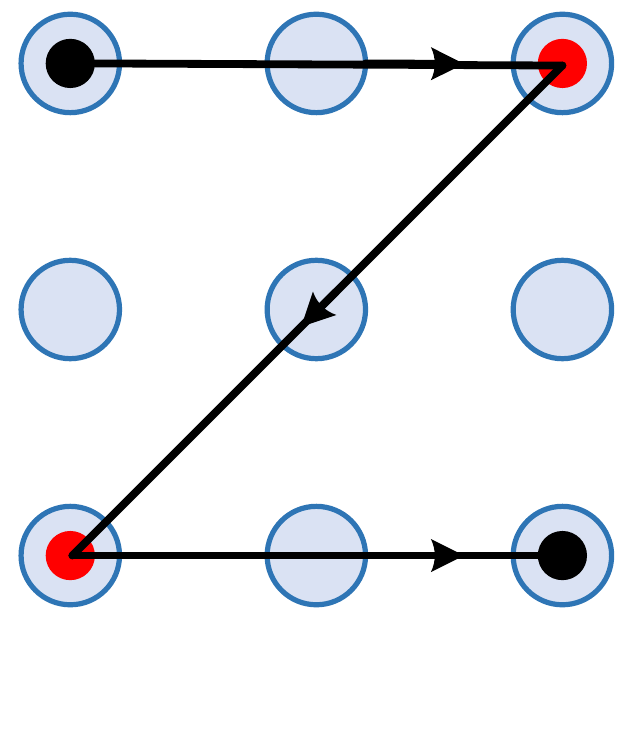}}
\subfigure[The corresponding C/O waveform]{
\includegraphics[width=0.59\columnwidth]{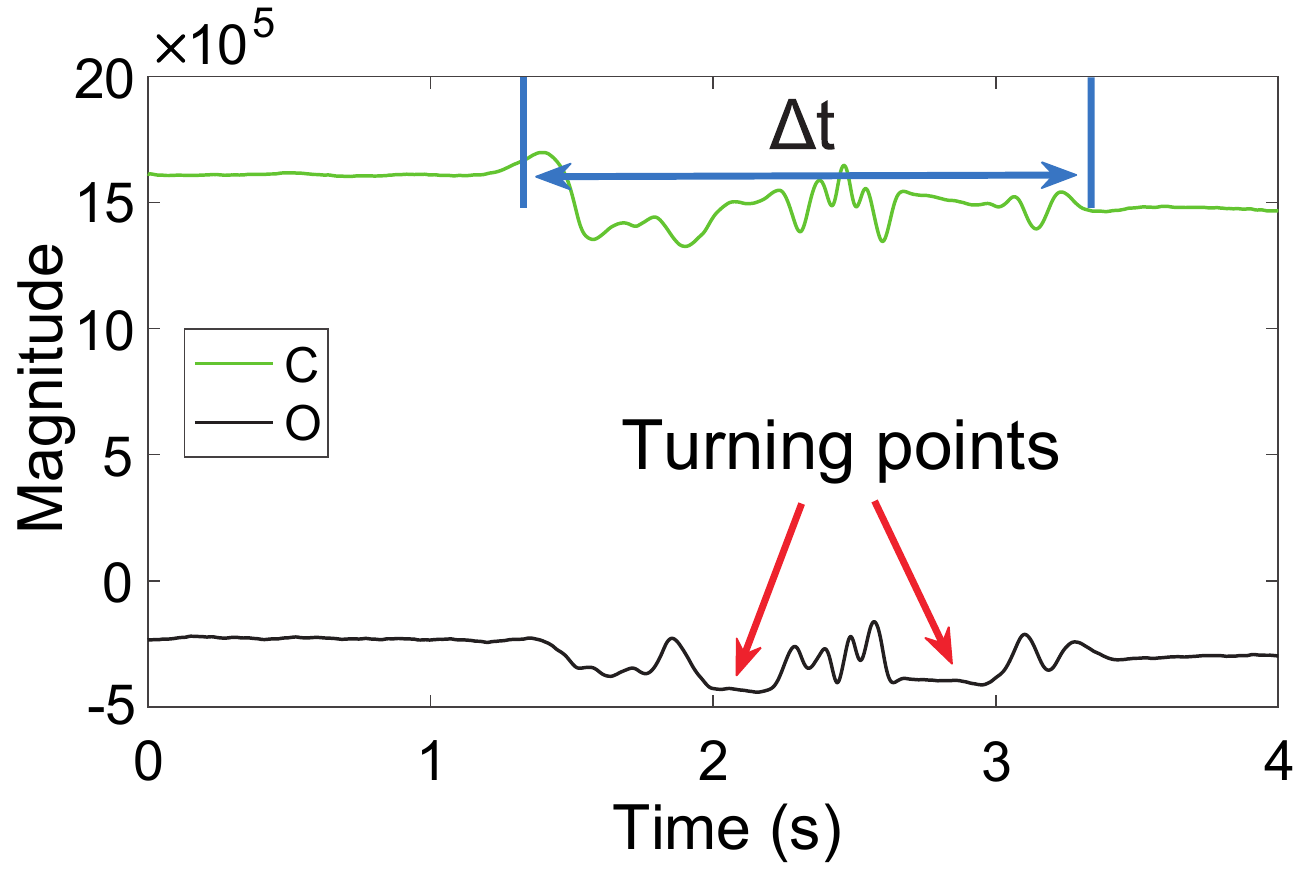}}
\caption{A fingertip draws a unlock pattern of ``Z'' during $\Delta t$. The red points in (a) are turning points, where the C/O waveform become relatively flat in (b).}
\label{fig:I-Q-wave}
\vspace{-3mm}
\end{figure}

Figure \ref{fig:I-Q-wave}(b) shows the C/O component corresponding to the unlock pattern in Figure \ref{fig:I-Q-wave}(a). The C and O components of the baseband signal have the same amplitude and frequency but \red2{different phases}. Because we use continuous wave acoustic signal with constant amplitude, the C/O waveform without fingertip motions \red2{is a flat line}. When the fingertip moves on the screen, the C/O waveform will fluctuate (e.g., the waveform within $\Delta t$ in Figure \ref{fig:I-Q-wave}(b) ).

\vspace{1mm}
\noindent{\textbf{Static Components Removal}:} The recorded acoustic signals are the combination of the true acoustic signals reflected by fingertip with the noisy acoustic signals. Most of the acoustic \red2{noise signals}, which travel through the \red2{line of sight (LOS)} path or are reflected by the surrounding objects, are the static components.
Therefore, we can remove the static components to obtain the true C/O components corresponding to the real signals reflected by the sliding fingertip.
To address this issue, we leverage the Local Extreme Value Detection (LEVD)~\cite{wang2016device} algorithm to estimate the static \red2{components}. 
We obtain the value of the static acoustic signal at the midpoint by computing the average value of two nearby maximum and minimum values, and leverage a linear interpolation algorithm to estimate the values of static acoustic signals on other points during fingertip movement.

Given the C/O waveform, once we find a local extreme point, we can compare it with the last extreme point. If their time interval is larger than an interval threshold $T_i$, it will be considered as a \red2{valid}  extreme point. Also, the local extreme point will be considered as an \red2{valid} extreme point only if their difference in amplitude is larger than the difference threshold $T_d$. The interval threshold $T_i$ is twice of the average time interval of two adjacent extreme points, which will be updated as more \red2{valid} extreme points are identified. The difference threshold $T_d$ is an empirical value that helps us to filter out local extreme points incurred by \red2{noises}.
}

\subsection{Signal Segmentation}\label{sec:LineSegment}

In order to reconstruct the unlock pattern, we first need to identify each line that is formed by the trajectory of the fingertip drawing on the screen. The signal segmentation phase is designed to segment the C/O component into fragments corresponding to each line of the pattern so that each line can be further identified. Note that, we can segment the signal manually or automatically. In PatternListener, we develop a {\color{black}\emph{Turning Points Identification} (TPI)} algorithm to realize automatic signal segmentation. Thereby, \red2{it} automatically infers unlock patterns of a large number of \red2{devices} simultaneously if the malware can collect signals from these users.

During the unlocking process, a new line starts when the fingertip makes a turn. For example, as shown in Figure~\ref{fig:I-Q-wave}(a), the fingertip turns twice for the unlock pattern of ``Z'' which consists of three separate lines. The point where the fingertip makes a turn is called a ``turning point'' (e.g., the two red points in Figure~\ref{fig:I-Q-wave}(a)). Thus, if we know the time of each turning point, we can segment the C/O component into fragments corresponding to each line.

\begin{figure}[!t]
\centering
\includegraphics[width=0.96\columnwidth]{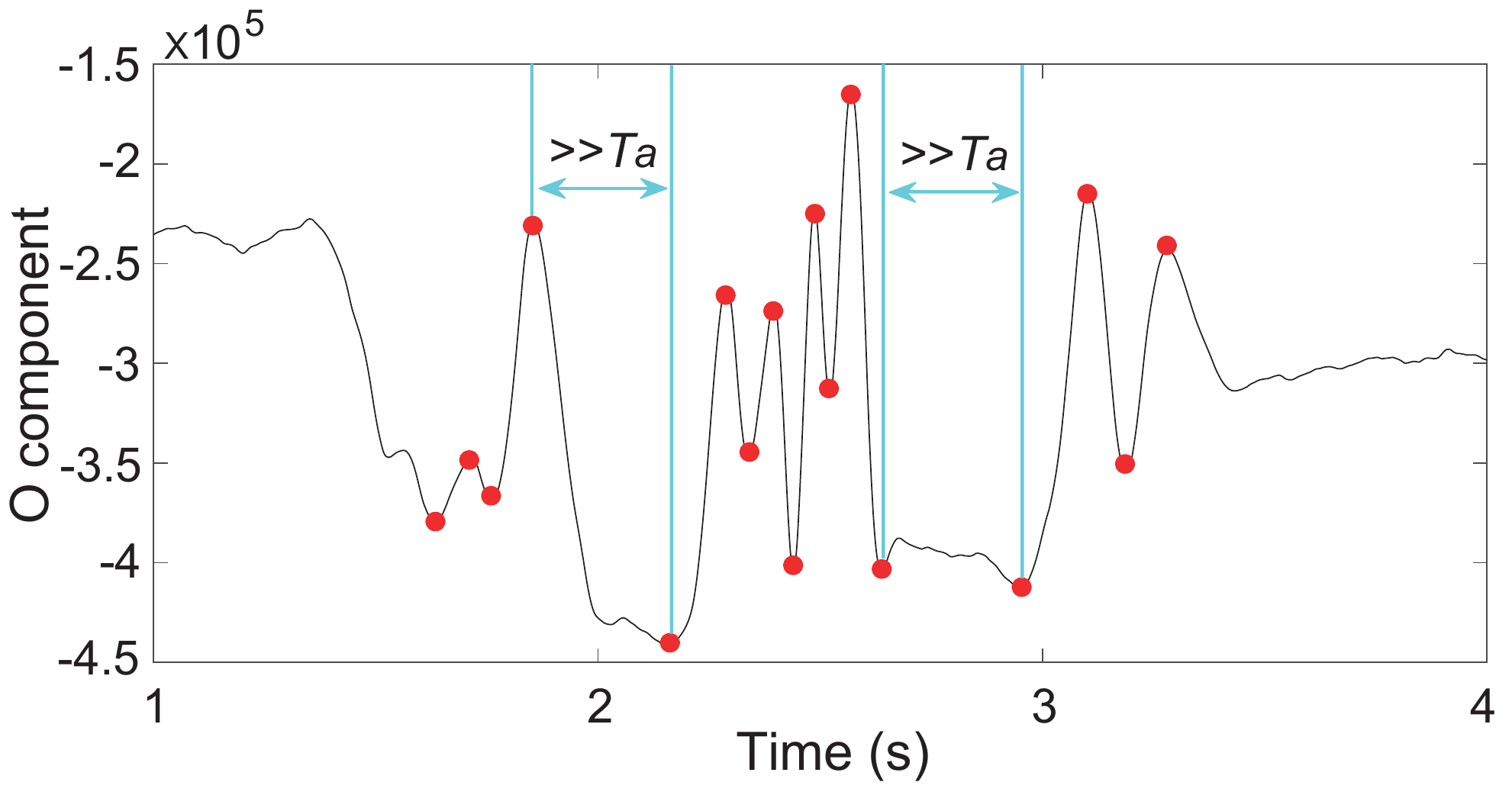}
\caption{An example of turning points identification. The red points are valid extreme points. $T_a$ is the average time interval between two adjacent extreme points.}
\label{fig:line-seg}
\vspace{-3mm}
\end{figure}

 \red2{Now we need }to identify the turning points of fingertip motions. \red2{We observe} that the fingertip pauses for a while (though the duration is very short) when arriving at a turning point. As a consequence, the acoustic signal reflected by the fingertip will be relatively \red2{stable} when the fingertip is at a turning point. That is, the C/O waveform fluctuates quickly when the fingertip moves normally and slowly at the turning points, as shown in Figure \ref{fig:I-Q-wave}(b).
Therefore, if we found that the time interval between two adjacent extreme points is much larger than the average time interval, this point is a turning point. Based on this observation, we propose a {\color{black}\emph{Turning Points Identification} (TPI)} algorithm to identify all the valid extreme points and further find the true turning points. Figure \ref{fig:line-seg} shows an example of turning points identification.


Note that, we have \red2{leveraged} the {\color{black} LEVD} algorithm to operate on the C and O component separately to find local extreme points (Section~\ref{sec:Preprocess}). However, some sharp noises which are introduced by environmental disturbance or hardware deficiency in C/O component may be identified as extreme points by the {\color{black} LEVD} algorithm \red2{mistakenly}. Considering that valid extreme points in C component are \red2{interlaced} in time with that in O component, we further sort the obtained extreme points of C/O component together according to time, exclude some misidentified extreme points to make the extreme points of C and O \red2{interlaced}. {\color{black}Finally, we can sequentially examine the time interval between two adjacent extreme points of C component to find all turning points.}

\begin{algorithm}[!t]\small{
\caption{ ~\textbf{TPI} algorithm}\label{alg:VEPI}
\LinesNumbered 
\KwIn{$C(t)$ and $O(t)$}
\KwOut{turning points $\{TP\}$ }
\vspace{1.9mm}
      /*Call the \textbf{LEVD} algorithm to operate on C/O component separately to get local extreme points*/\\
      $ \{LE_C\}\gets \textbf{LEVD}(C(t))$\;
      $ \{LE_O\}\gets \textbf{LEVD}(O(t))$\;
      /*Exclude some misidentified extreme points to make the extreme points of C and O interlaced in time*/\\
      $\{LE_{CO}\}\gets TimeSort(\{LE_C\},\{LE_O\})$\;
      $(\{EC\},\{EO\})\gets Alternate(\{LE_{CO}\})$\;
    {\color{black}   $n\gets Num(\{EC\})$\;
      /*Calculate the average interval of the extreme points of C*/\\
      $In_{ave}\gets AveInter(\{EC\})$\;
      /*Sequentially examine the time interval between two adjacent extreme points of C to find all turning points*/\\
      \For{$i = 2$ to $n$}{
      $In_i\gets Interval(EC_i)$\;
      \If{$In_i >> In_{ave}$}{
      $\{TP\}\gets EC_i$\;
      }
      }
      }
}
\end{algorithm}

\subsection{Relative Movement Measurement}\label{sec:Measurement}

After the signal fragments corresponding to each line are accurately segmented, {\color{black} we identify and re-evaluate startpoints and endpoints of the C/O components and then} measure the relative movement of the fingertip associated with each line. 

\begin{figure}[!t]
\centering
\subfigure[C/O waveform over time]{
\includegraphics[width=0.5\columnwidth]{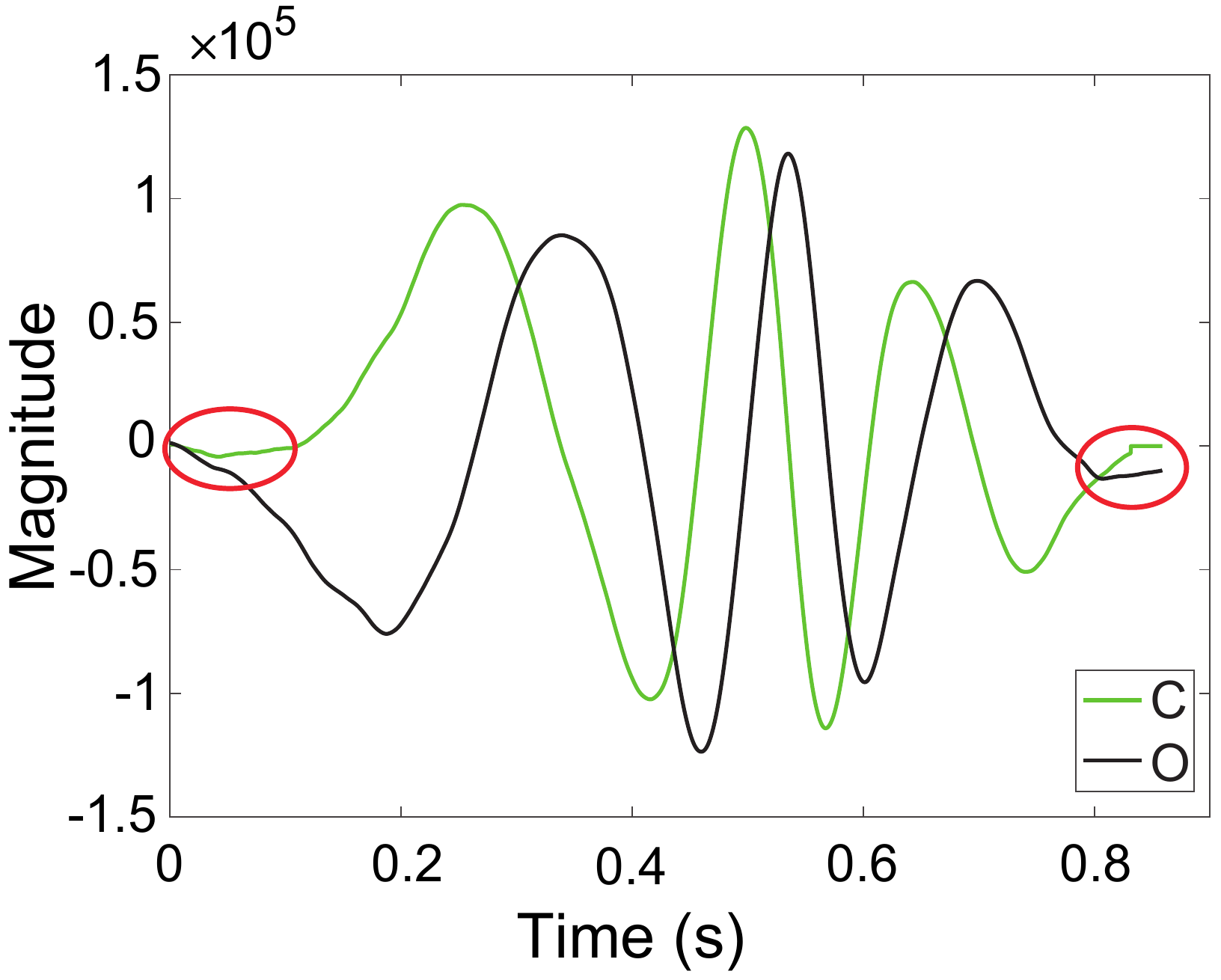}}
\subfigure[C/O trace]{
\includegraphics[width=0.41\columnwidth]{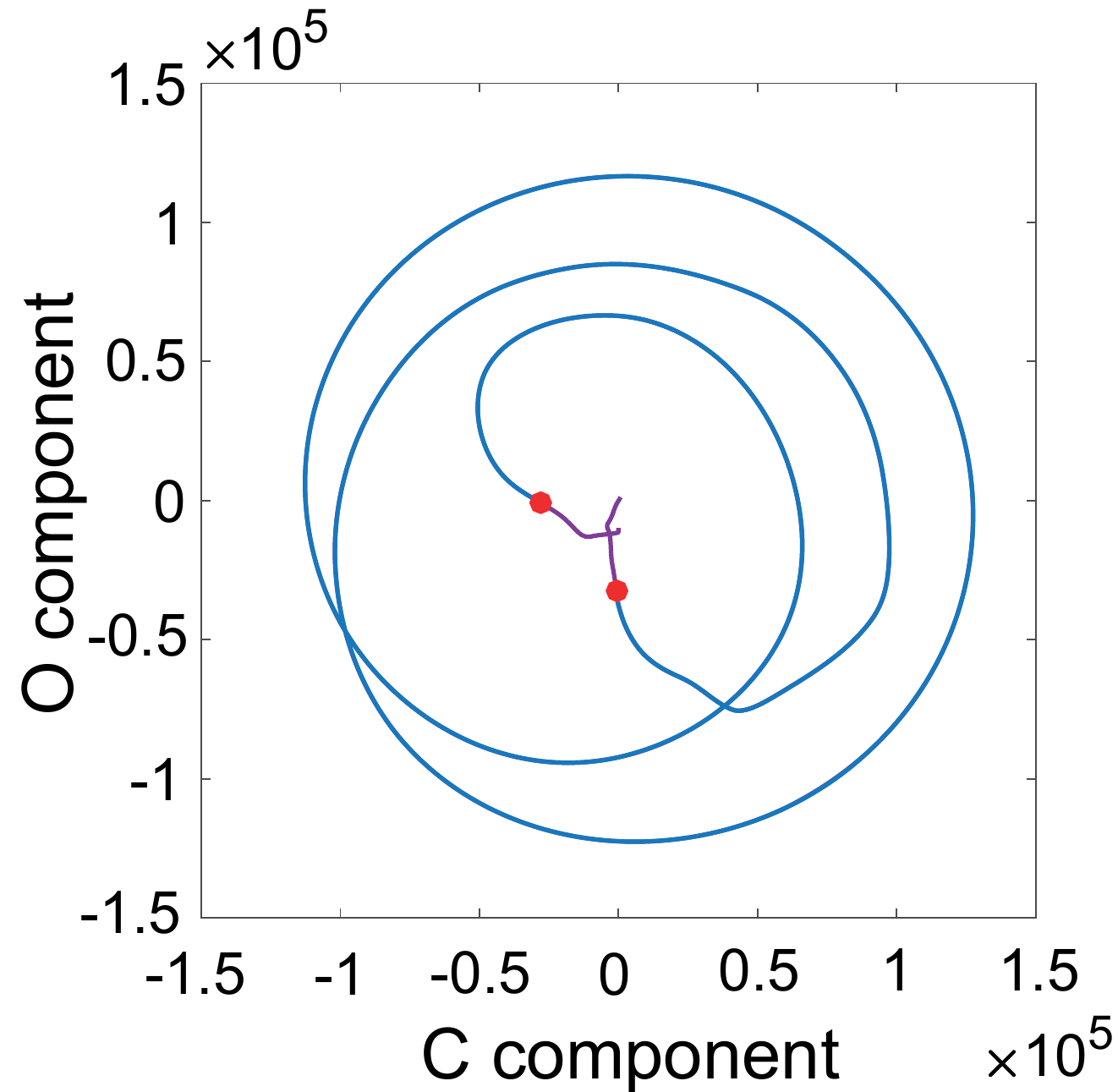}}
\caption{The C/O component corresponds to the one reflected by the moving fingertip.}
\label{fig:I/Q_moving}
\vspace{-3mm}
\end{figure}

\vspace{1mm}
\noindent{\textbf{Startpoint and Endpoint Re-identification}:}
Figure \ref{fig:I/Q_moving} shows the C/O component corresponding to a pattern line after removing noises. The C/O waveform approximates a sinusoid and the C/O trace is similar to a circle whose center is $(0,0)$. However, the identification of \red2{the} C/O component's startpoint and endpoint is not very accurate due to the error of signal segmentation. \red2{An example of identification error is shown in the area of the ellipse} in Figure \ref{fig:I/Q_moving}(a). Now we will describe how to accurately calculate the phase changes of acoustic signals due to fingertip movement for each line and reduce the identification error.

\begin{figure}[!t]
\centering
\includegraphics[width=0.6\columnwidth]{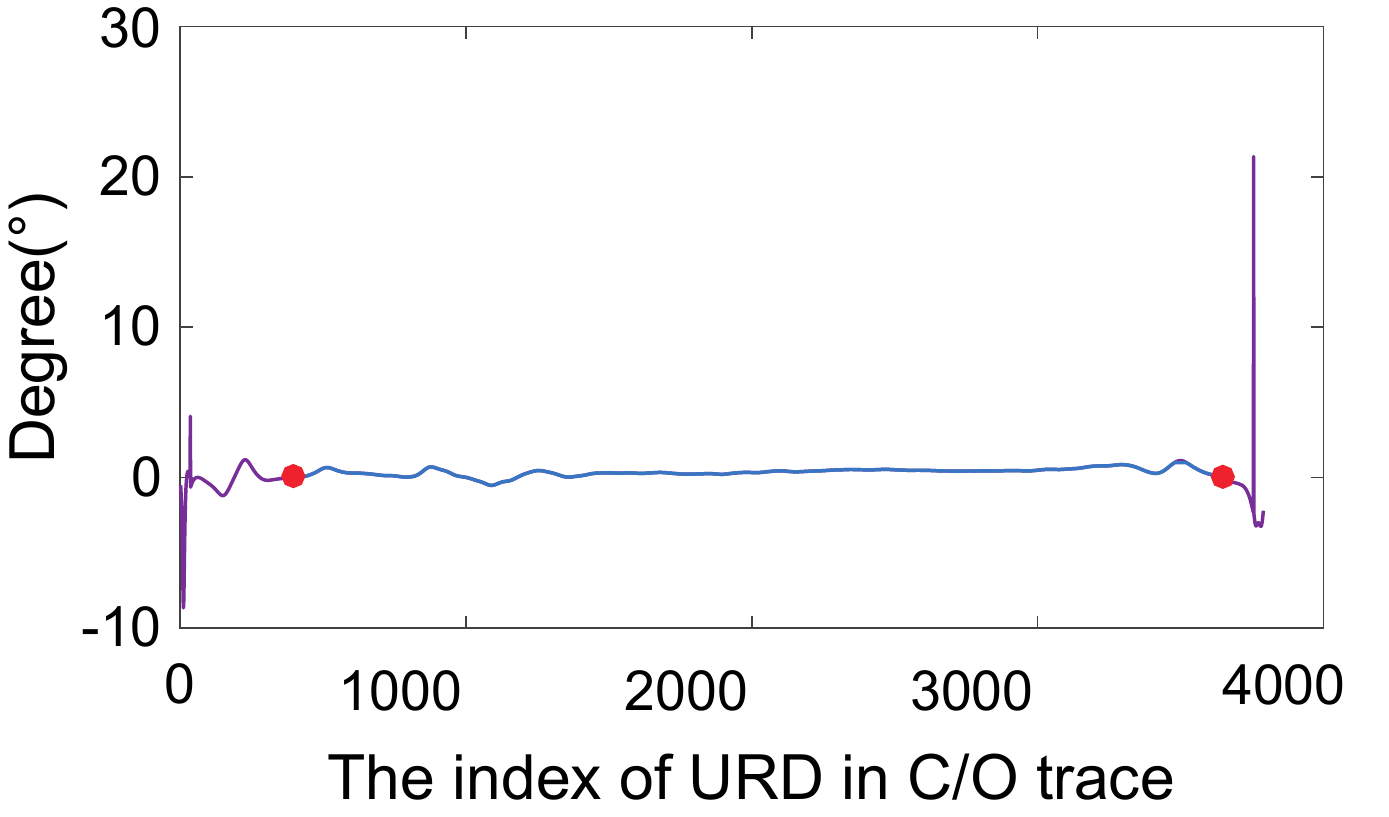}
\caption{The URDs in C/O trace. The two red points correspond to the more  accurate startpoint and endpoint of C/O trace.}
\label{fig:URD}
\vspace{-3mm}
\end{figure}
\vspace{1mm}

The basic idea is to calculate the accumulated rotating degrees of all points in C/O trace. The points in C/O trace are denoted as $P_1$, $P_2$, ..., $P_i$, ..., $P_n$, where $i$ is the time index. The rotating degree of the arc $\wideparen{P_1P_2}$ can be approximatively calculated as the angle between line $P_1P_2$ and line $P_2P_3$. We use $URD$ to denote the rotating degree of adjacent two points. Similarly, we can obtain the rotating degree of arc $\wideparen{P_2P_3}$, ..., $\wideparen{P_{n-2}P_{n-1}}$. Finally, the phase change of a fragment of acoustic signals is equal to the sum of all $URD$ in its C/O trace. Figure \ref{fig:URD} shows the URDs in a C/O trace. We can see that the values of most URDs in this C/O trace are about $0.6^{\circ}$, but some URDs at the beginning and ending period have unstable values. In fact, the value of URD is proportional to the sliding speed of \red2{the} fingertip. Since the sliding speed of the fingertip is relative stable, the points in the C/O trace whose URDs vary greatly and \red2{deviate} from  normal values are invalid. According to this observation, we can find the more accurate startpoint and endpoint for the C/O trace (e.g., the two red points in Figure \ref{fig:I/Q_moving}(b)) and further calculate the phase changes more \red2{accurately}.

\noindent{\textbf{Relative Movement of the Fingertip}:}
In PatternListener, we leverage the phase-based approach~\cite{wang2016device} to measure the relative movement of the fingertip by calculating changes of the phases of the acoustic signals reflected by the fingertip, and then convert the changes of the phase into the changes of path lengths. Note that, the fingertip movement will affect both the frequency and the phase of reflected signals. Here, the frequency is influenced by the movement speed, while the phase is impacted by the movement distance and direction. The movement distance and direction for the same pattern will not change, while the movement speed may vary. Therefore, we use the phase-based approach rather than Doppler shift-based approach~\cite{chen2014airlink} \red2{that} extracted movement features \red2{corresponding to} the frequency changes and \red2{is} significantly impacted by the movement speed.


Let $d(t)$ denote the path length of acoustic signals \red2{reflected by the moving fingertip} at time $t$, $\phi(t)$ denote the phase of acoustic signals reflected by \red2{the} fingertip at time $t$, and $\lambda$ denote the wavelength of acoustic signals. Then the path length change during the time period $(t_1, t_2)$ can be calculated as \red2{follows:}
\begin{equation}\label{eq:PathLength}
d(t_2)-d(t_1) = \frac{-\lambda}{2\pi}(\phi(t_2)-\phi(t_1)).
\end{equation}
\red2{According to }Equation \ref{eq:PathLength}, we can \red2{obtain} the path length change during any time period, which is determined by the relative movement of the fingertip. Given that the speed of sound $\nu$ in the air is 340m/s and frequency of acoustic signal $f$ is 19 kHz, we can obtain the wavelength $\lambda = \nu/f$ is 1.79 cm. Therefore, the phase based distance measurement approach is enough to distinguish different fingertip movements on the pattern grid.

\subsection{{Pattern Line Inference}}

Now we use the path length changes to infer each line \red2{constituting} the unlock pattern. We first characterize the movement feature related to the sliding fingertip and then infer the line with a similarity measurement.

\begin{figure}[!t]
\centering
\includegraphics[width=0.7\columnwidth]{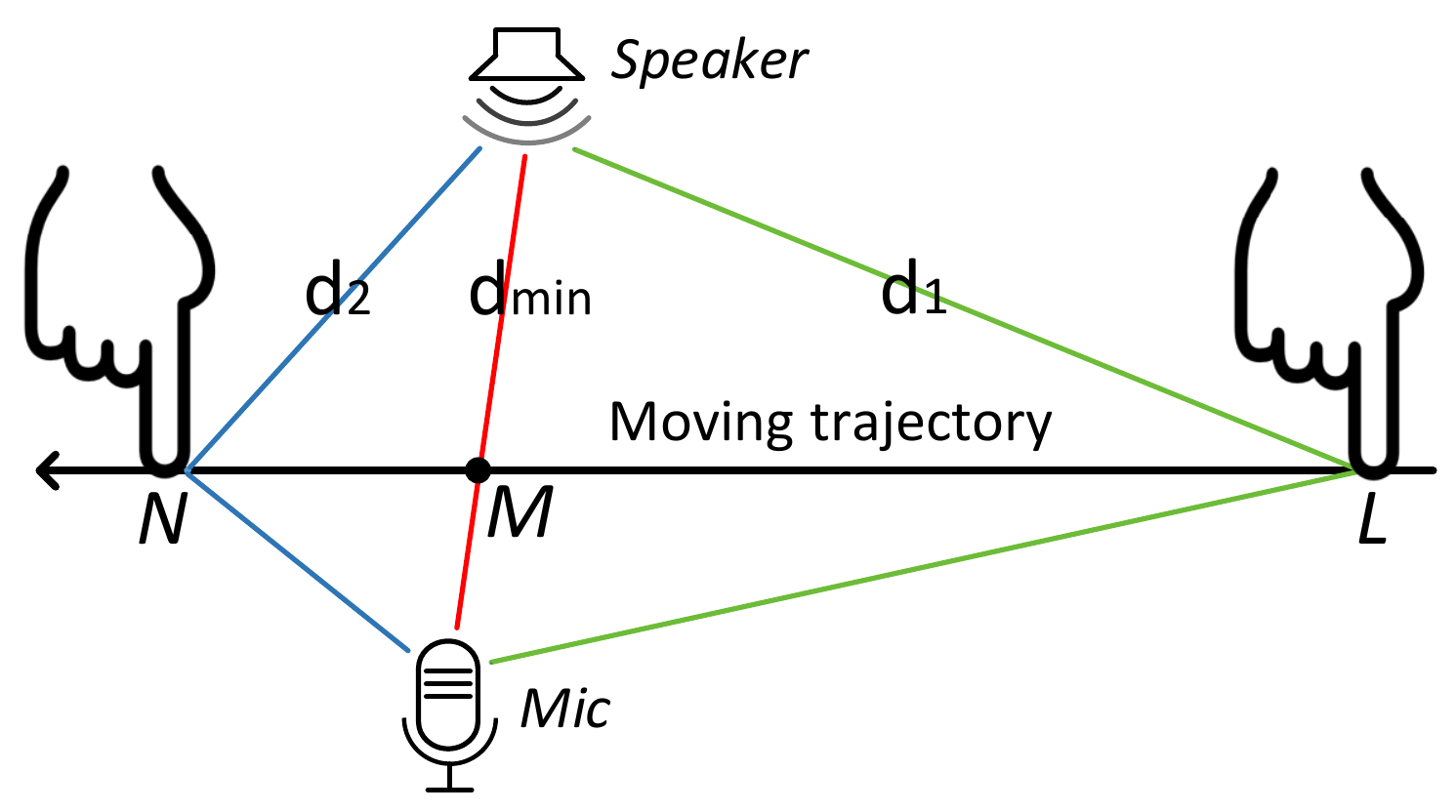}
\caption{The speaker and microphone are on different sides of the  trajectory of the fingertip.}
\label{fig:different}
\vspace{-3mm}
\end{figure}

\subsubsection{Movement Feature Extraction}

In fact, the relationship between the changes of the path lengths and relative movement of the fingertip \red2{is decided by} the positions of the speaker and the microphone. Without loss of generality, we consider two situations: the speaker and the microphone are at different sides of the trajectory of the fingertip, and the speaker and the microphone are at the same side of the  trajectory. It is worth noting that the trajectory of the fingertip is a line because we have segmented the sound signal into fragments corresponding to each line.

\vspace{1mm}
\noindent{\textbf{Different Sides}:} The path length changes of acoustic signals reflected by the sliding fingertip  are shown in Figure \ref{fig:different}.
A trajectory of the fingertip is a line between the speaker and the microphone with an arbitrary length and direction.
We assume that the fingertip slides from $L$ to $N$ and passes $M$, where $M$ is the intersection between the trajectory and the line from the speaker to the microphone. As we know, the path length of the acoustic signal will decrease from $L$ to $M$ and then increase from $M$ to $N$. In other words, the path length changes only have three cases: always increasing, always decreasing, and \red2{increasing after decreasing}.
Therefore, we can use a two-dimensional vector $(d1,d2)$ as the fingertip movement feature, \red2{where} $d1$ is the path length changes from $L$ to $M$, and $d2$ is the path length changes from $M$ to $N$. 

\begin{figure}[!t]
\centering
\includegraphics[width=0.7\columnwidth]{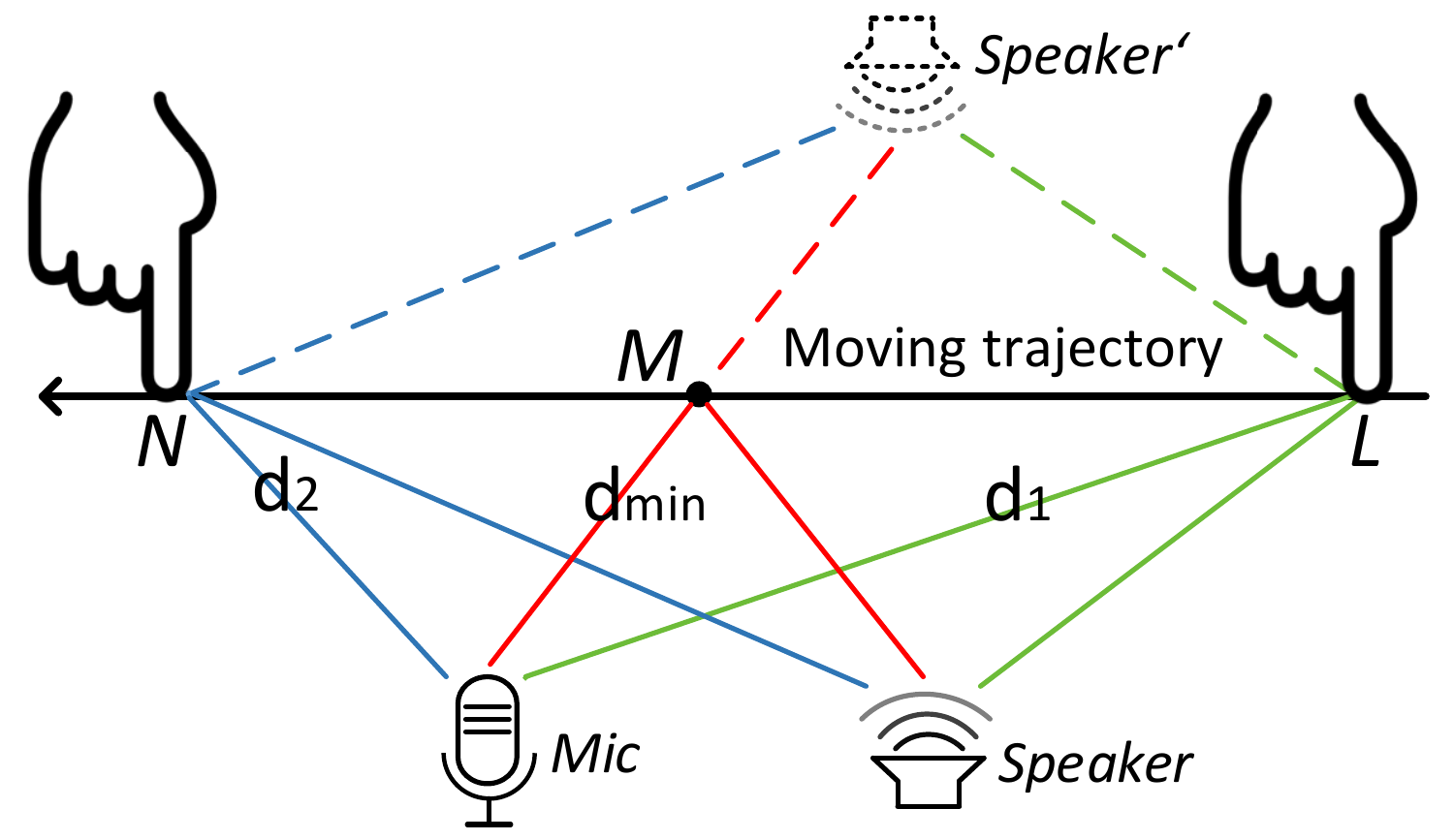}
\caption{The speaker and microphone are on the same side of the trajectory of the fingertip.}
\label{fig:same}
\vspace{-3mm}
\end{figure}

\vspace{1mm}
\noindent{\textbf{The Same Side}:} The path length changes of acoustic signals reflected by the moving fingertip are shown in Figure \ref{fig:same}. We cannot directly observe how the path length of acoustic signal changes from $L$ to $N$.  To solve the problem, we assume there exists a virtual speaker $Speaker'$, which is the mirrored speaker along the  trajectory. Thus, the path length of acoustic signals \red2{between} the speaker \red2{and} the moving fingertip is always the same as that \red2{between} $Speaker'$ \red2{and} the moving fingertip. That is, the path length changes from the speaker reflected by the moving fingertip to the microphone are always the same as that from $Speaker'$ reflected by the moving fingertip to the microphone. Therefore, the path length change in this case is similar to that the speaker and the microphone are at different sides of the trajectory of the fingertip. Hence, we can still use a two-dimensional vector $(d1,d2)$ as the {fingertip movement feature}.

There are more than one speaker and microphone in most commercial off-the-shelf mobile devices. PatternListener requires at least one pair of speaker and microphone to infer the unlock pattern. \red2{The attack effectiveness} will be better if more speakers and microphones are used. The fingertip feature varies  \red2{with} different pairs of speaker and microphone because the changes of the path length are directly \red2{impacted} by the positions of the speaker and the microphone. Therefore, we can combine the two-dimensional feature vectors of different pairs of speaker and microphone to infer each line more accurately. To prevent the interference among acoustic signals \red2{generated} from different speakers, \red2{signals generated from different speakers can use different frequencies.}

\subsubsection{Similarity based Line Inference}

We build a ground-truth database of feature vectors for each line with different pairs of speaker and microphone. 
Given a start point, the fingertip may slide to other 8 points in the 3 $\times$ 3 grid to make up 8 different lines. We compare the feature vector 
with that of 8 different lines and calculate the corresponding similarity. Let $(d1_{ij},d2_{ij})$ denote the feature vector of the $i_{th}$ ($i\in[1,8]$) line with the $j_{th}$ pair of speaker and microphone, $(d1'_j, d2'_j)$ denote the extracted feature vector with the $j_{th}$ pair of speaker and microphone. The similarity $S_{ij}$ between {the extracted feature vector} and that of the $i_{th}$ line with the $j_{th}$ pair of speaker and microphone is calculated as \red2{follows:}

\begin{equation}\label{eq:Similarity1}
S_{ij} = 1- \frac{\sqrt{(d1_{ij}-d1'_j)^2 + (d2_{ij}-d2'_j)^2}}{\sqrt{(d1_{ij})^2+(d2_{ij})^2} + \sqrt{(d1'_j)^2+(d2'_j)^2}}.
\end{equation}

Then, we combine the feature vectors with different pairs of speaker and microphone to \red2{obtain} the similarity $S_{i}$ between {extracted feature vectors} and that of the $i_{th}$ line:

\begin{equation}\label{eq:Similarity2}
S_{i} = \sum_{j=1}^{n} W_j S_{ij}, \hspace{3mm}  W_1+W_2+...+W_n=1,
\end{equation}
where $n$ is the total number of pairs of speaker and microphone, $W_j$ is the weight coefficient of the $j_{th}$ pair of speaker and microphone. We set different weight coefficients for different pairs of speaker and microphone since the relative movement measurement result is usually more reliable when the place of the pair of speaker and microphone is closer to the sliding fingertip. When the similarity $S_{i}$ is larger than a threshold $T_s$ (Here, we empirically set $T_s=0.65$ according to our measurement results), we will treat the $i_{th}$ line with the start point as a candidate line. Note that, there may exist multiple candidates for the current line. We need to enumerate all these candidates.

\begin{figure}[!t]
\centering
\subfigure[The unlock pattern]{
\includegraphics[width=0.36\columnwidth]{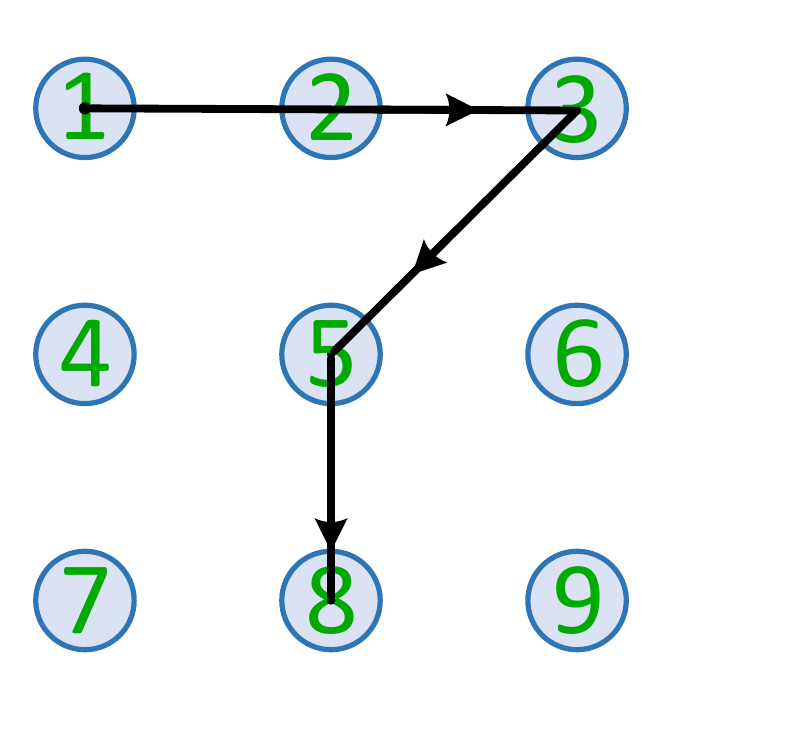}}
\subfigure[Pattern tree]{
\includegraphics[width=0.4\columnwidth]{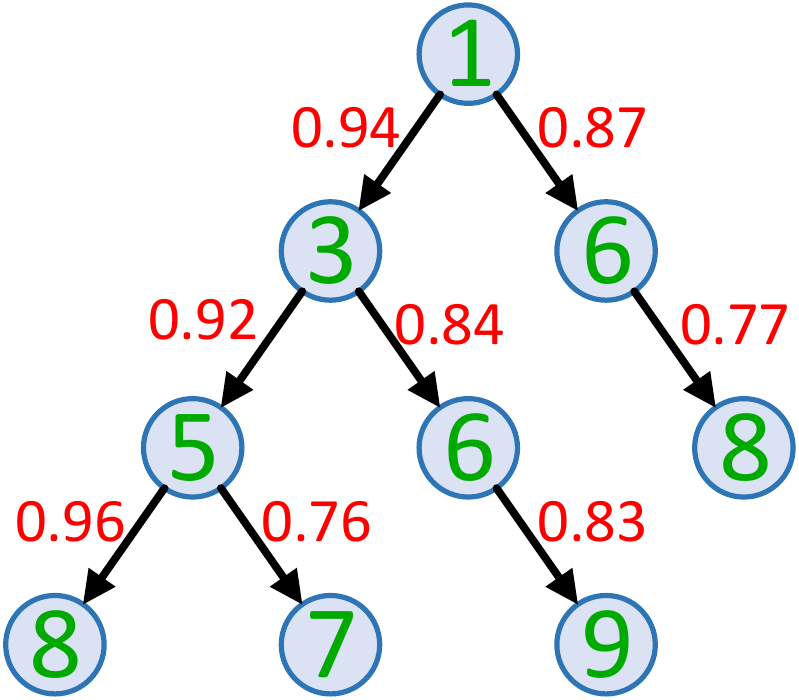}}
\caption{The process of generating candidate patterns. The red number in (b) is the corresponding similarity.}
\label{fig:Pattern_generating}
\vspace{-3mm}
\end{figure}

\subsection{Candidate Patterns Generation}\label{sec:Generating}

We map all candidates for the line into the pattern grid to generate the pattern after the pattern line is identified. We propose a pattern tree to \red2{conduct} the pattern reconstruction and filter out impossible candidates according to the fact that the lines of the pattern are sequentially connected. At last, the top-$5$ patterns with the highest similarities are selected as the candidates for the unlock pattern.

Figure \ref{fig:Pattern_generating}(a) shows a $3 \times 3$ grid and we \red2{name} the dots from $1$ to $9$. With the candidates for each line, we use the \red2{multiway tree} to build a pattern tree (as shown in Figure \ref{fig:Pattern_generating}(b)) to generate candidates for the unlock pattern. The root of the tree is the start point of the pattern. If multiple start points are inferred for the pattern, we generate multiple pattern trees correspondingly {so that we can estimate the start point when the fingertip starts clicking on the screen according to the data generated by the motion sensors.} For a pattern tree, we will add the candidates of the first line at the first layer, the candidates of the second line at the second layer, and so on until the candidates of the last line are added into the tree. The weight of each edge \red2{indicates} the corresponding similarity of the candidate. For example, the candidates for the line $1\rightarrow 3$ are $1\rightarrow 3$ and $1\rightarrow 6$, so they are added at the first layer and their similarities are added as the weights.

We can \red2{conclude} that each path from the root to a leaf at the last layer is a {candidate pattern}. \red2{As shown in Figure \ref{fig:Pattern_generating}(b)}, $1\rightarrow 3\rightarrow 5\rightarrow 8$, $1\rightarrow 3\rightarrow 5\rightarrow 7$, and $1\rightarrow 3\rightarrow 6\rightarrow 9$ are the candidates of the unlock pattern. Note that some branches of the tree do not reach the last layer (e.g., $1\rightarrow 6\rightarrow 8$), which is because no suitable candidate can be found after the previous line.
The similarity of a {candidate pattern} 
can be defined as the average similarity of all lines on the path. 
{The higher the average similarity on the path is, the more likely it is that the path corresponds to the actual unlock pattern.} We calculate the similarities of all paths for all pattern trees and rank them from high to low. The top five paths/patterns with the highest similarities will be considered as the candidates for the unlock pattern.



%% file: evaluation.tex
\begin{figure}[!t]
\centering
\includegraphics[width=0.9\columnwidth]{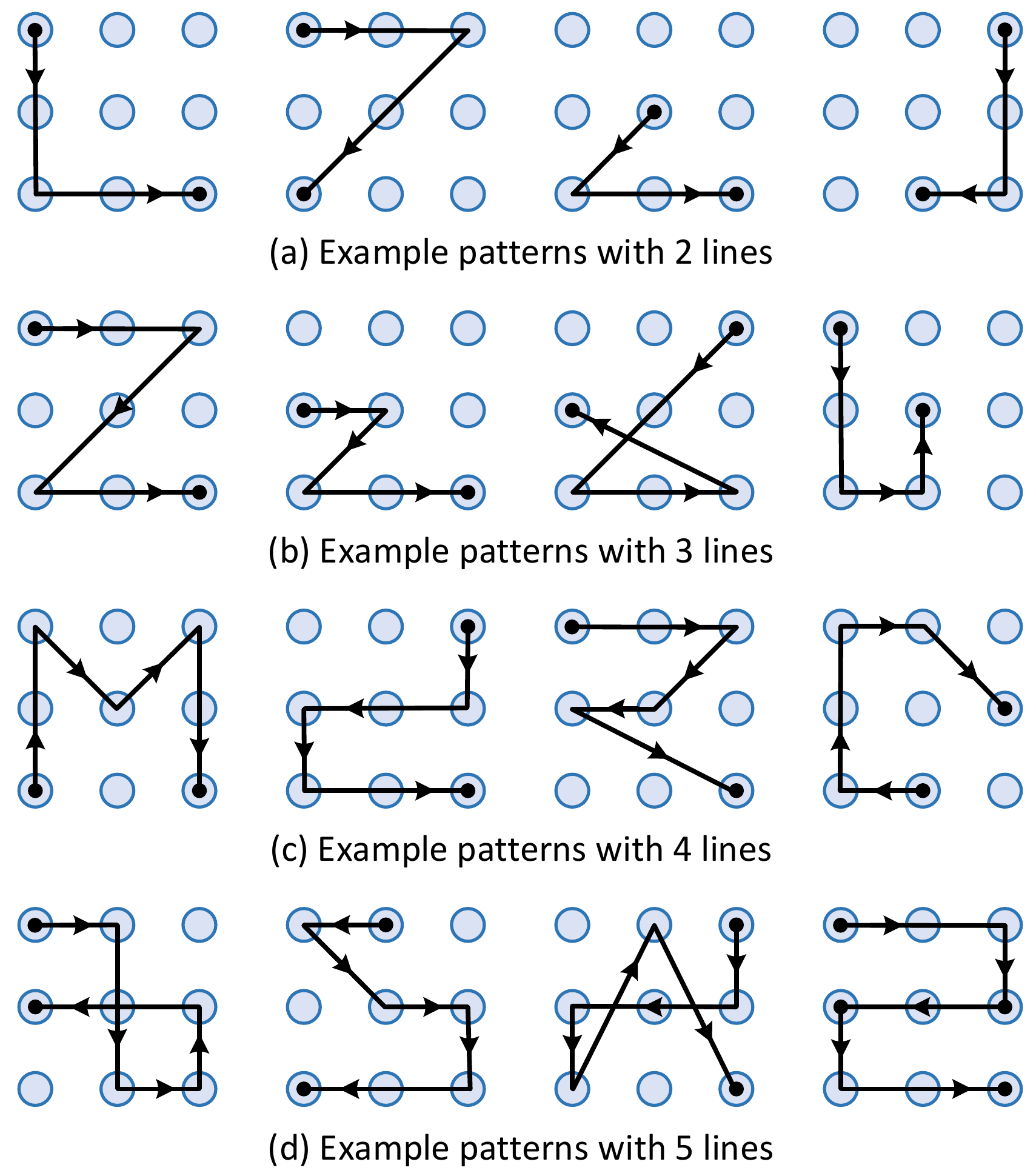}
\caption{Example patterns with different number of lines.}
\label{fig:pattern-line}
\end{figure}

\section{System Evaluation}\label{sec:evaluation}

In this section, we present our experimental results based on our PatternListern prototype \red2{installed} on off-the-shelf smartphones.

\subsection{Experimental Setup}

\subsubsection{Experiment Setup}
We implement a PatternListener \red2{app and install it} on off-the-shelf smartphones. {As we discussed in Section~\ref{sec:overview}, 
PatternListener can be disguised as a benign APP and run in the background once it is installed.} We evaluate PatternListener on two different smartphone platforms: SAMSUNG C9 Pro and HUAWEI P9 Plus. The server is a PC with 2.9 GHz CPU and 8 GB memory. Note that there are more than one speaker and microphone in most mobile devices. However, most smartphones have the issue of hardware echo cancellation when under dual track recording, which will affect the feature extraction from different pairs of speaker-microphone. Therefore, we only use one microphone to record the acoustic signals reflected by the fingertip.

\subsubsection{Ground-truth Construction}
In PatternListener, the fingertip movement features are impacted by the relative positions of the speaker and the microphone.
{ Since the relative positions of speakers and microphones are usually identical in the same smartphone model but may differ in different smartphone models, we only need to generate the ground-truth database of  the  features according to each smartphone model rather than each smartphone device.
In addition,} in order to construct the ground-truth database, we only need to extract the features of all lines rather than the features of all patterns. The total number of all possible lines in a $3\times3$ pattern grid is only 72 ($=9\times8$), while that of all possible patterns is $389,112$~\cite{uellenbeck2013quantifying}.
Therefore, it is not difficult to  build the ground-truth database to validate PatternListener.

\begin{figure*}[!th]
\begin{minipage}[t]{0.32\linewidth}
\centering
\includegraphics[width=0.9\textwidth]{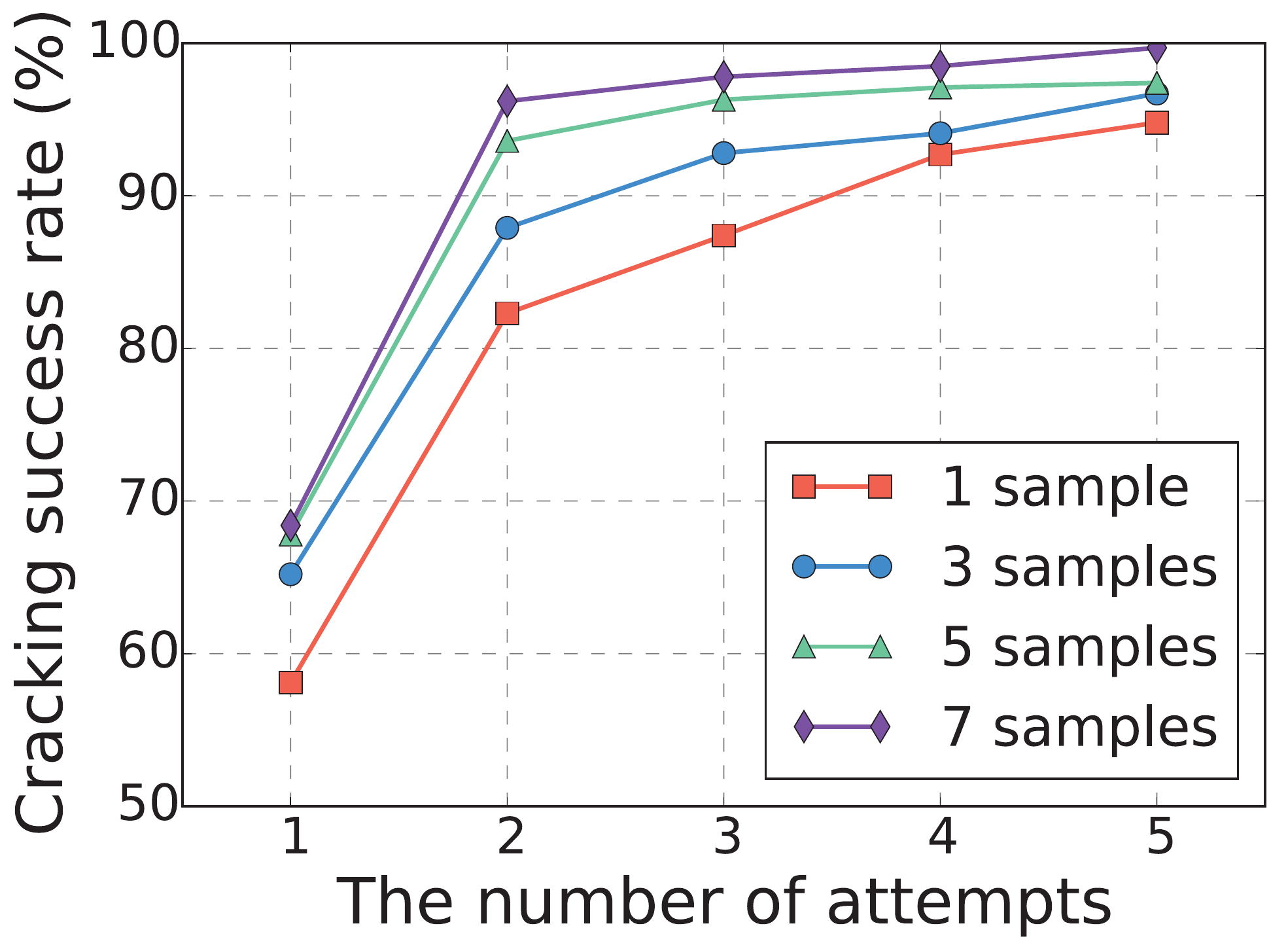}
\caption{Overall cracking rate.}
\label{fig:Overall-seccess}
\end{minipage}
\begin{minipage}[t]{0.32\textwidth}
\centering
\includegraphics[width=0.9\columnwidth]{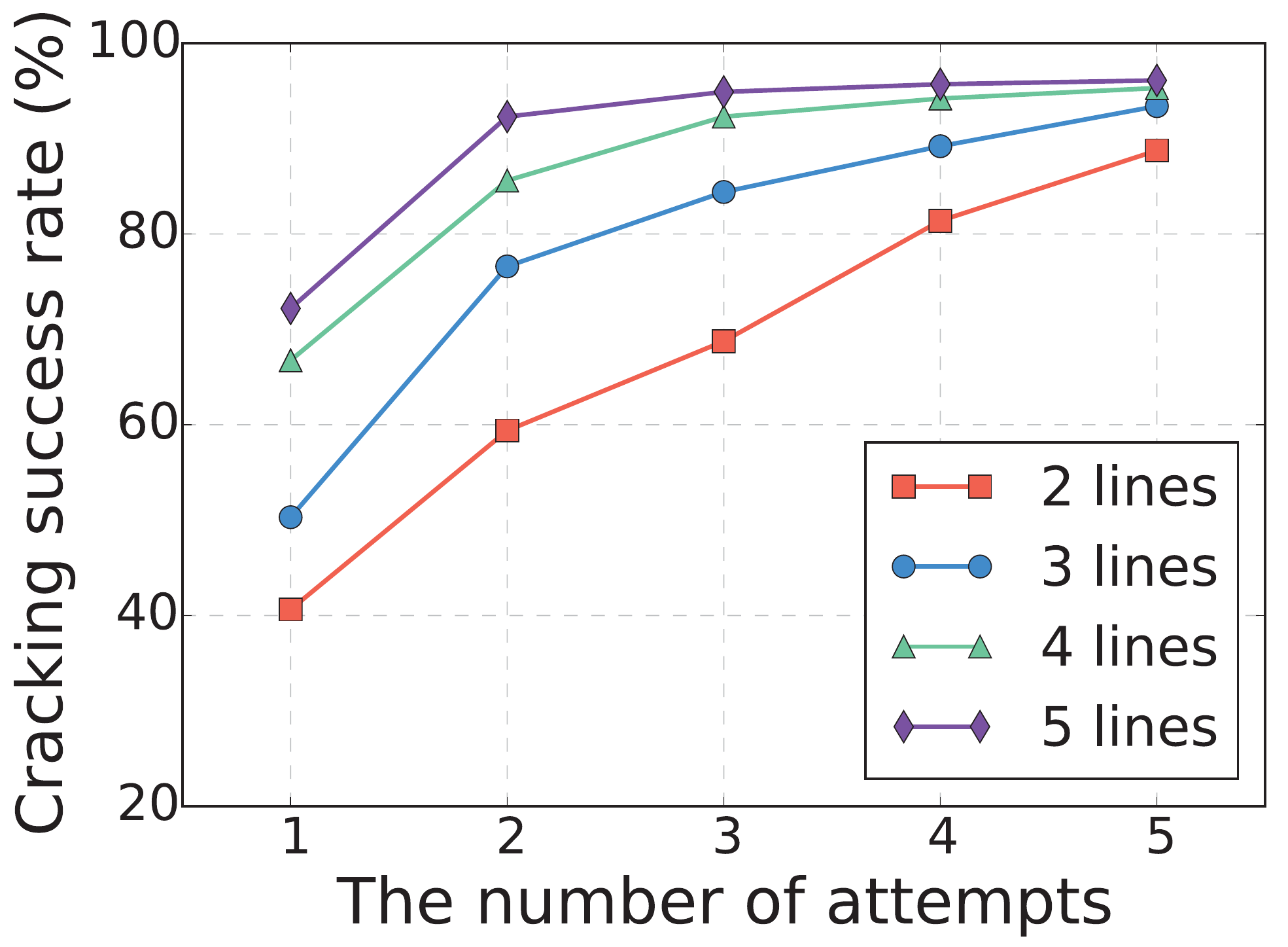}
\caption{Impact of pattern complexity.}
\label{fig:Complexity-seccess}
\end{minipage}
\begin{minipage}[t]{0.32\textwidth}
\centering
\includegraphics[width=0.9\columnwidth]{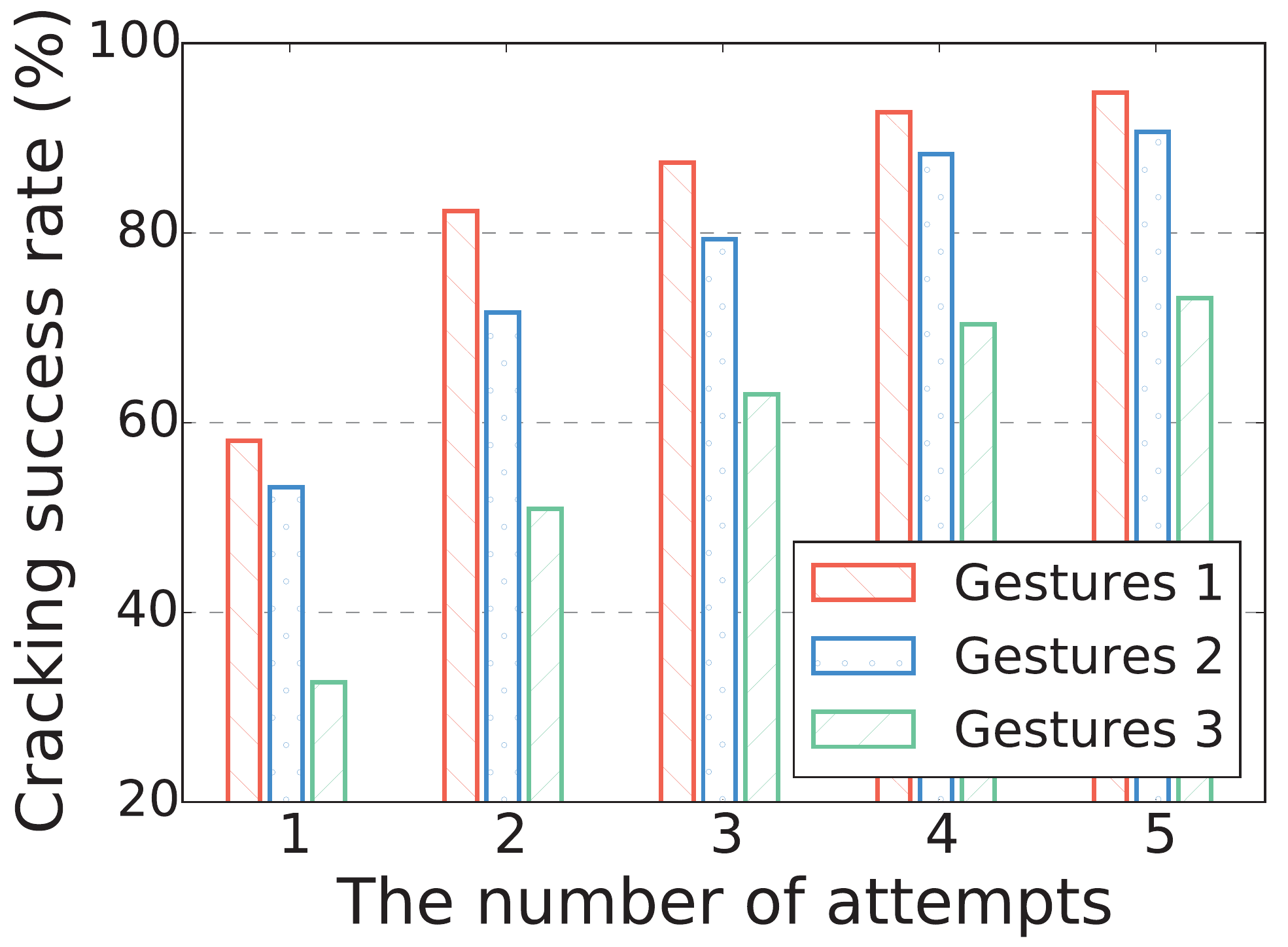}
\caption{Impact of gesture.}
\label{fig:Gestures-seccess}
\end{minipage}
\end{figure*}

\subsubsection{Data Collection}
{In order to collect the data of various patterns, we generated 500 anonymous questionnaires \red2{for} volunteers who are using or have used the pattern lock and collected 197 unique patterns. 
To evaluate the influence of various patterns on the accuracy of PatternListener, we selected 120 typical patterns, i.e., 30 patterns with 2 lines, 30 patterns with 3 lines, 30 patterns with 4 lines, and 30 patterns with 5 lines. Figure~\ref{fig:pattern-line} shows the example patterns with different numbers of lines we used in experiments.} We found that a large number of people (about $38\%$) start the pattern from the top left-most point of the pattern grid. 
\red2{Besides,} we add another 10 patterns \red2{with} familiar alphanumeric characters, since a recent report showed that people tend to set these patterns as the unlock pattern owing to their preference to familiar pictures~\cite{loge2015tell}.


\subsubsection{Default Setting}
\begin{figure}[t!]
    \centering
    \subfigure[Gesture 1]{\includegraphics[width=0.3\linewidth]{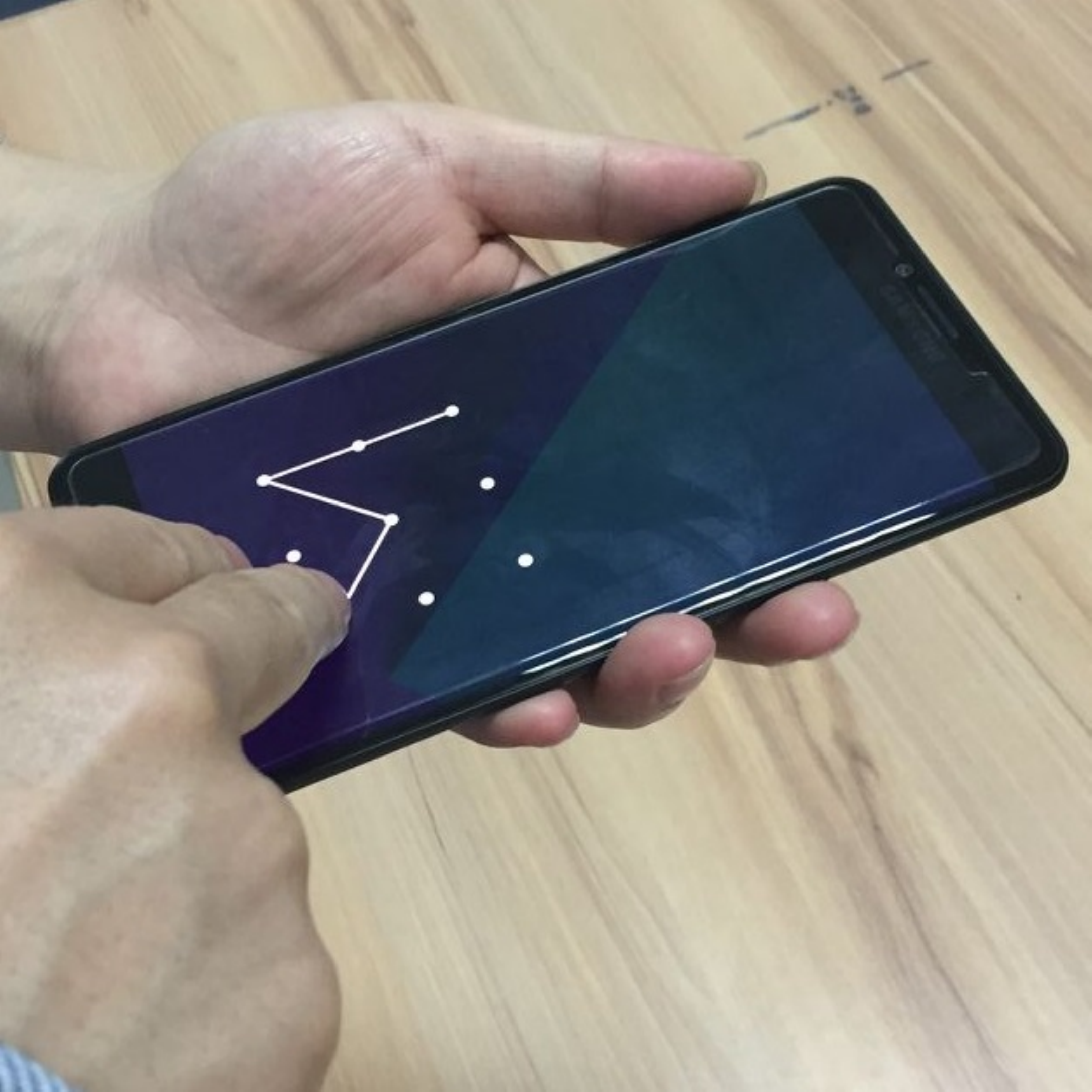}}
    \subfigure[Gesture 2]{\includegraphics[width=0.3\linewidth]{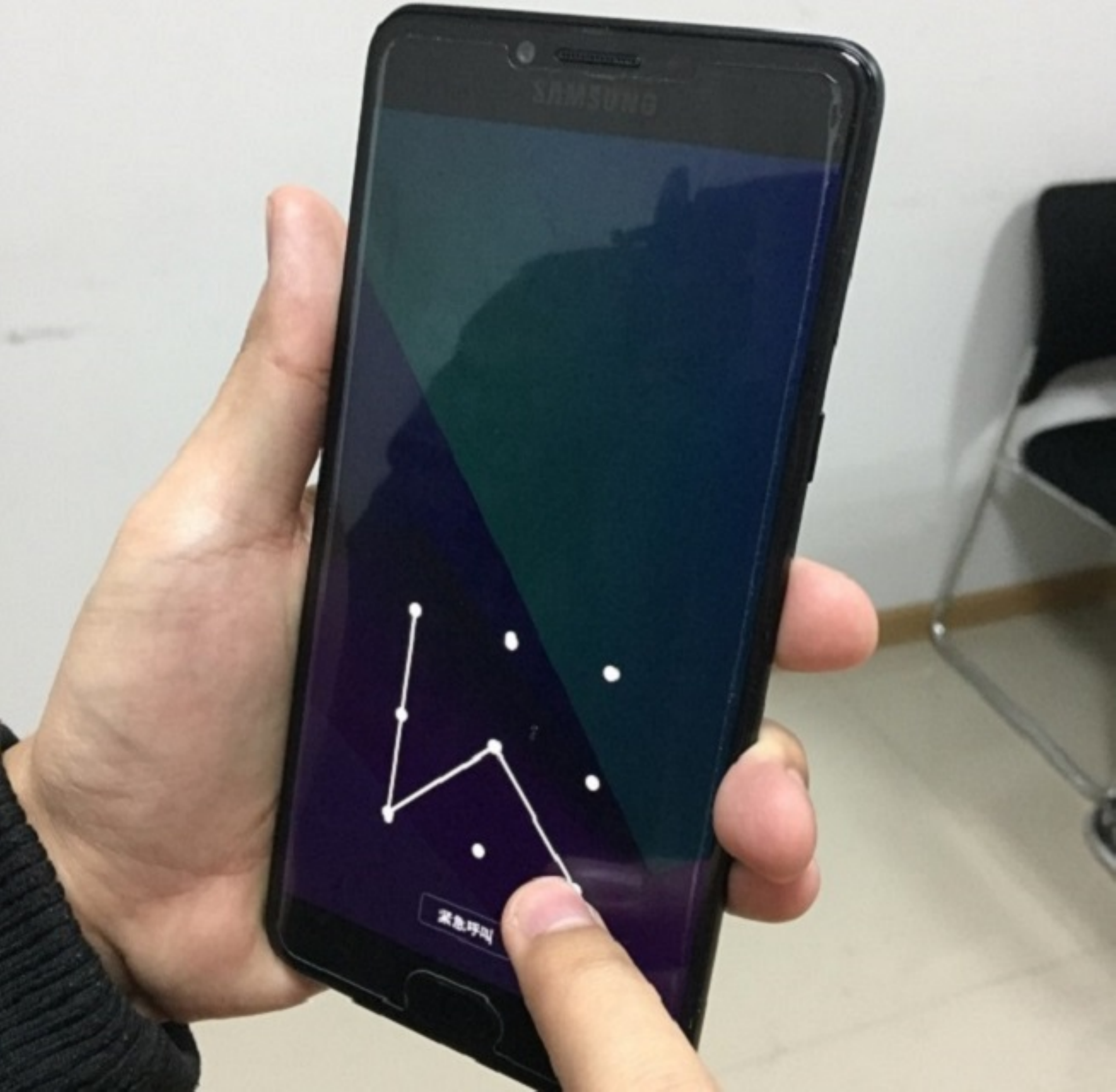}}
    \subfigure[Gesture 3]{\includegraphics[width=0.3\linewidth]{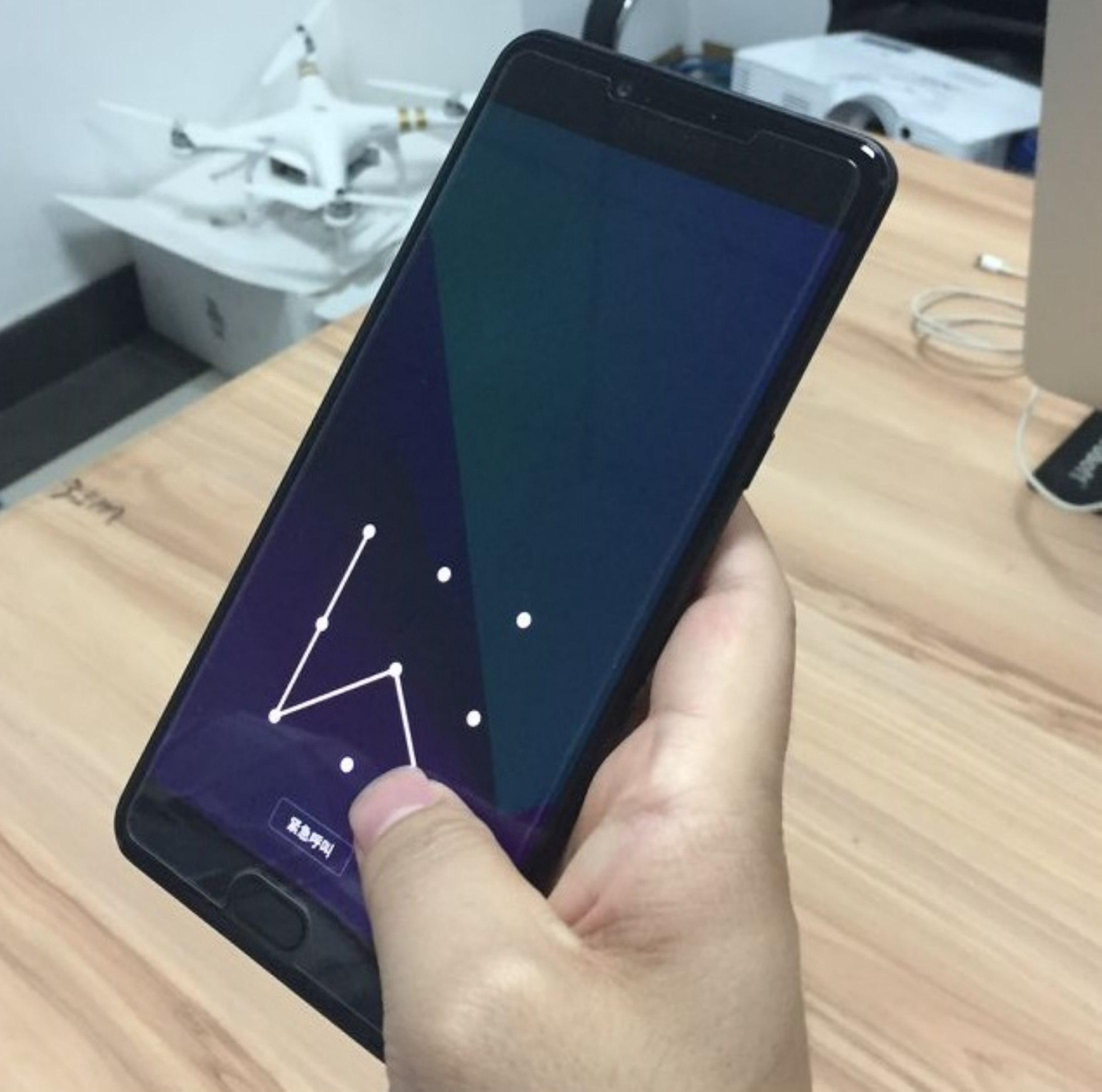}}
    \caption{Different unlock gestures.}
    \label{fig:unlock-gesture}
\end{figure}

We recruited 5 volunteers, i.e., three males and two females, to reproduce the {\color{black}130 collected patterns independent of their own unlock patterns} on the Android $3\times3$ default pattern grid of two target smartphones: a SAMSUNG C9 Pro running Android 6.0.1 and a HUAWEI P9 Plus running Android 7.0. {The generated acoustic signals are at the range of $18 \sim 20$ kHz with multiple frequencies since the measurements obtained from different frequencies can be combined to \red2{improve accuracy}, and the sampling rate of the microphone is 48 KHz. 
\red2{Note that, the key difference of drawing patterns among different individuals is the drawing speed.
Actually, the difference of success rates between various drawing speeds is below 10\% (see Figure~\ref{fig:Speed-seccess}). Thus, the experiment results with more people are similar.} 
The Android system allows at most 20 consecutive failed unlock attempts, and the device will be temporally locked for 30 seconds after five failed attempts. Thus, we evaluate the success rate for inferring patterns within five attempts. In most of our experiments, we used SAMSUNG C9 Pro as the evaluation platform and drawn the unlock patterns with a moderate speed {\color{black}(i.e., the usual drawing speed of each participant)} when the smartphone is horizontally held with a hand (i.e., Gesture 1 shown in Figure~\ref{fig:unlock-gesture}) in an office.}

\subsection{Experimental Results}

\subsubsection{Overall Success Rate}
We first present the overall  success rate of cracking patterns with different numbers of samples of  the 130 collected patterns. It is feasible to infer the same unlock pattern with multiple samples since PatternListener can run in the background for a long time and capture various samples of the same pattern, which help to improve the success rate. {\color{black}Here, a sample means a piece of acoustic signal corresponding to one unlock process. }Figure~\ref{fig:Overall-seccess} shows the overall success rate with one to five attempts. First of all, PatternListener achieves an average success rate of $58.1\%$ with only 1 sample at the first attempt. The success rate will increase to $94.8\%$ with five attempt, which is a very exciting result. Since the Android system allows up to five failed attempts before temporally locking the device, we can conclude that PatternListener can successfully crack most pattern locks in practice. In addition, the success rate will increase with more samples since the influence of noisy samples will be eliminated. Specifically, the success rate of five attempts reaches $99.7\%$ with 7 samples. Therefore, PatternListener is very effective and accurate at reconstructing unlock patterns.

\subsubsection{Impact of Pattern Complexity}
This experiment evaluates the influence of pattern complexity to PatternListener, which aims to validate if more lines included in a pattern can provide stronger security. Note that, we define the complexity of a pattern by the number of lines instead of the existing metrics~\cite{sun2014dissecting}, which is decided by the number of points and intersections. The reason is that PatternListener considers  a combination of {several individual}
lines that are sequentially connected as a pattern. Figure~\ref{fig:Complexity-seccess} demonstrates the success rate with only 1 sample under different pattern complexities. We can observe that the cracking success rate becomes higher for more complicated patterns, which is an interesting finding that contradicts people's intuition. {The reason is that the patterns with more lines also contain more fingertip movement features. It also validates the effectiveness of the proposed pattern tree that can remove more irrelevant candidates if more lines are set within a pattern. Therefore, the complicated pattern with more lines cannot provide stronger protection if under the attack of PatternListener.}  

\begin{figure*}[!th]
\begin{minipage}[t]{0.32\linewidth}
\centering
\includegraphics[width=0.9\textwidth]{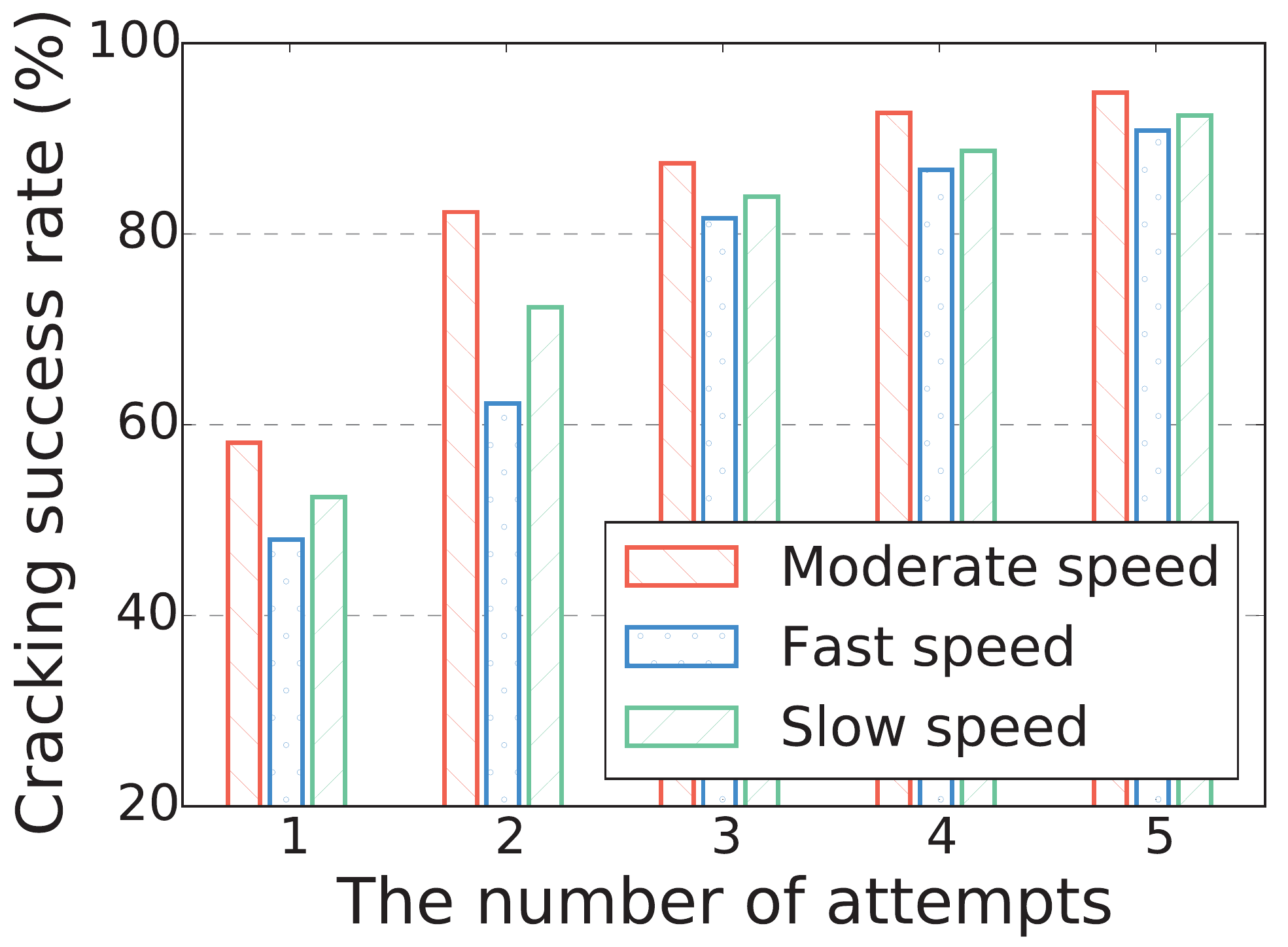}
\caption{Impact of speed.}
\label{fig:Speed-seccess}
\end{minipage}
\begin{minipage}[t]{0.32\textwidth}
\centering
\includegraphics[width=0.9\columnwidth]{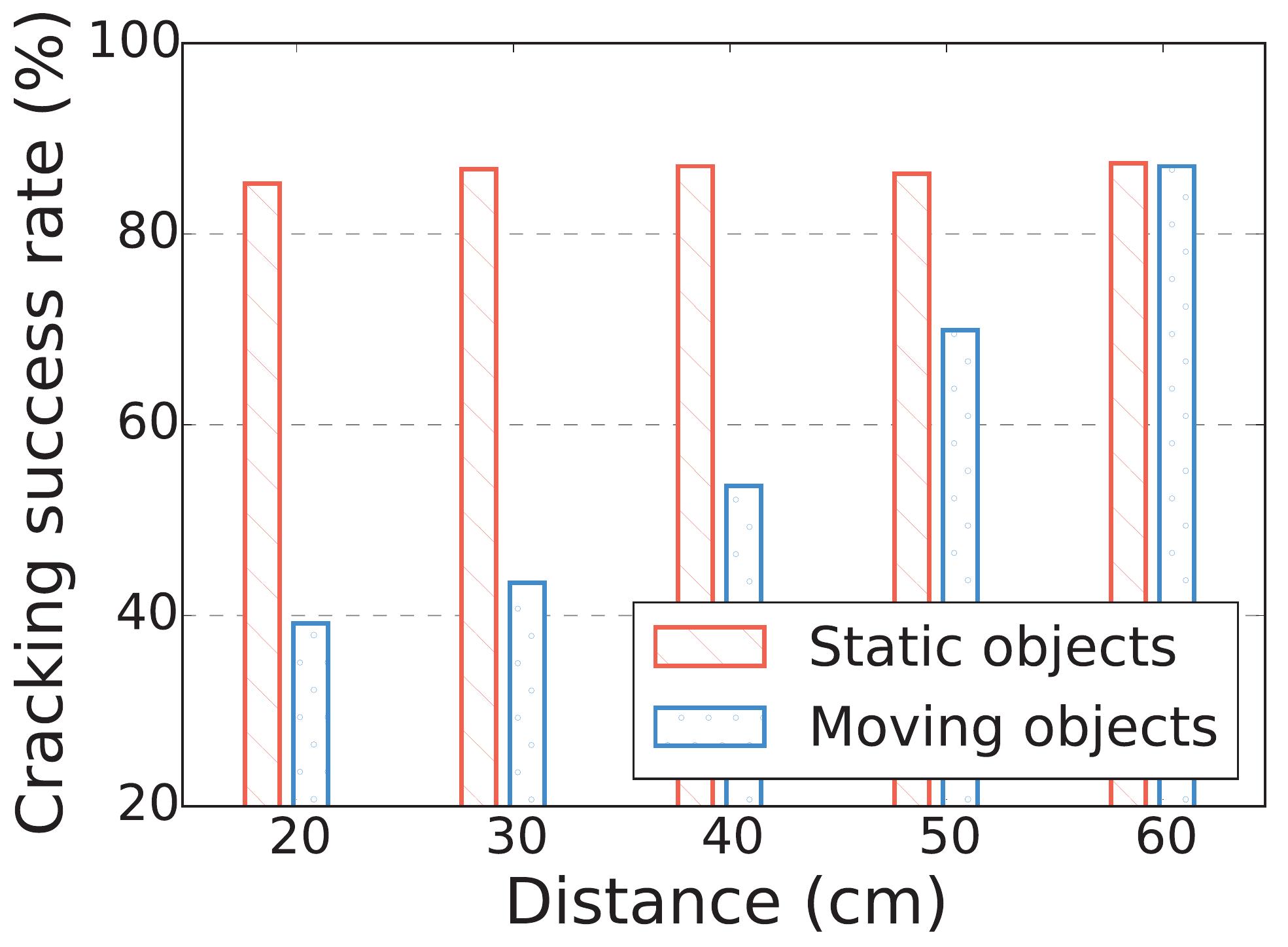}
\caption{Impact of surrounding objects.}
\label{fig:Objects-seccess}
\end{minipage}
\begin{minipage}[t]{0.32\textwidth}
\centering
\includegraphics[width=0.9\columnwidth]{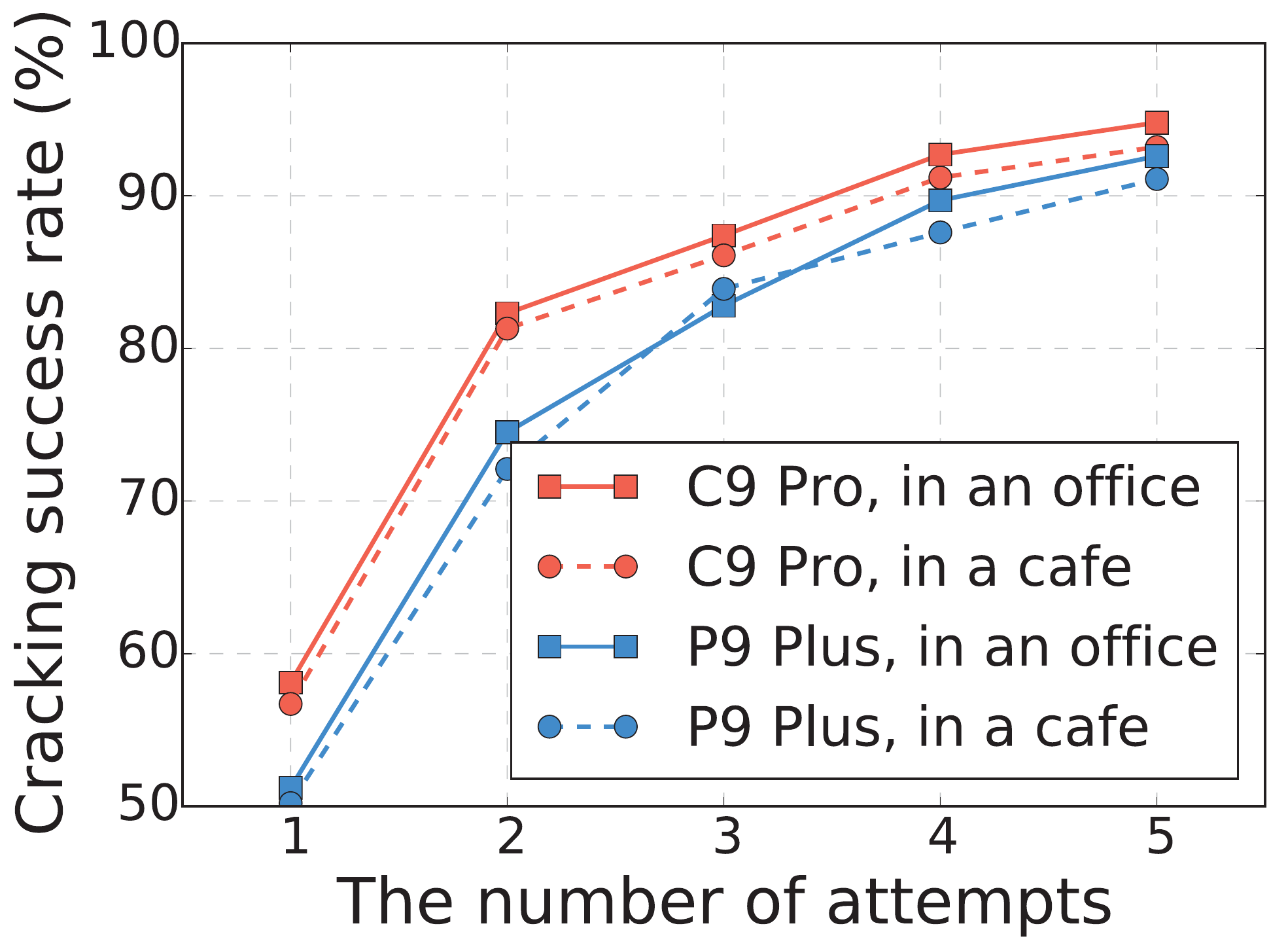}
\caption{Impact of smartphone models and noise.}
\label{fig:Smartphone-seccess}
\end{minipage}
\end{figure*}

\subsubsection{Impact of Gesture}
We then investigate the influence of gesture on the accuracy of pattern cracking. Since people have their own habits to hold and unlock their phones, as shown in Figure~\ref{fig:unlock-gesture}, we should ensure that PatternListener can infer the unlock pattern accurately under various holding gestures: {(i) in gesture 1, the victim holds the phone with one hand horizontally and draws the pattern with another hand, (ii) in gesture 2, the victim holds the phone with one hand vertically and draws the pattern with another hand, } (iii) in gesture 3, the victim holds and unlocks the phone with the same hand (i.e., holding the phone with the right hand and drawing a pattern with the thumb).

Figure~\ref{fig:Gestures-seccess} demonstrates the success rate with 1 sample under different gestures. We can observe that PatternListener achieves the best accuracy (average success rate is 94.8\% in five attempts) for gesture 1, while the worst (average success rate is 73.2\% in five attempts) for gesture 3. {The reason is that the ground-truth database is only constructed with gesture 1. There is more movement noise generated by {\color{black}the one-handed operation in gesture 3}. However, according to Figure~\ref{fig:Overall-seccess}, we can know that the success rate in the worst situation can be increased by capturing more samples. Note that the accuracy with gesture 2 is not significantly lower than that with gesture 1, which means that PatternListener is robust to the change of device orientation.} Hence, we can conclude that PatternListener is relatively robust under different holding gestures.


\subsubsection{Impact of Drawing Speed}
We further study the impact of drawing speed on the success rate of PatternListener. Even though people usually draw the pattern at a moderate speed to avoid mistakenly connecting the wrong dots, the preferred speed of individual still varies. Hence it is worth figuring out the range of drawing speed that PatternListener can support so as to guarantee that it can stay robust to the changes of drawing speed. To collect the data with different drawing speeds, we ask the participants to draw patterns with different speeds, i.e., moving the fingertip moderately, quickly, or slowly.

Figure~\ref{fig:Speed-seccess} demonstrates the success rate with 1 sample under different drawing speeds. We discover that a speed that is too fast or too slow can exert a slight negative influence on the pattern inference. Patterns with a moderate speed have the highest average success rate. For the patterns that are drawn too fast, the audio signal segmentation is not very accurate. While for the patterns that are drawn slowly, the prolonged recording process introduces accumulated errors to our algorithm. However, the difference under different speeds decreases with more attempts.
In addition, the speed of drawing patterns in practice will not vary as extremely as that in our experiments. Therefore, PatternListener is relatively robust to the changes of drawing speed, especially with more attempts.

{
\subsubsection{Impact of Surrounding Objects}
In this experiment, we investigate the influence of the surrounding objects that interference with the acoustic signals on the accuracy of pattern cracking. We perform this experiment with two participants, i.e., a participant drawing patterns while another participant's hand acting as the surrounding object at different distances away from the phone. Figure~\ref{fig:Objects-seccess} shows the success rate of one sample with three attempts while the background hand is static or keeps moving at various distances. In PatternListener, the cracking success rate is almost not affected by the static objects because the \textit{Static Components Removal} can remove the noisy acoustic signals reflected the surrounding static objects. We can observe that the surrounding moving objects obviously affect the success rate, however, the effect decreases as the distance increases. When the distance exceeds 60 cm,  the influence of surrounding objects becomes negligible. This is because the acoustic power decays as 2 times of the square of the distance from the phone to the surrounding objects. This experimental result demonstrates that it is not easy to disrupt PatternListener by using surrounding objects.
}

\begin{figure}[!t]
\centering
\includegraphics[width=0.8\columnwidth]{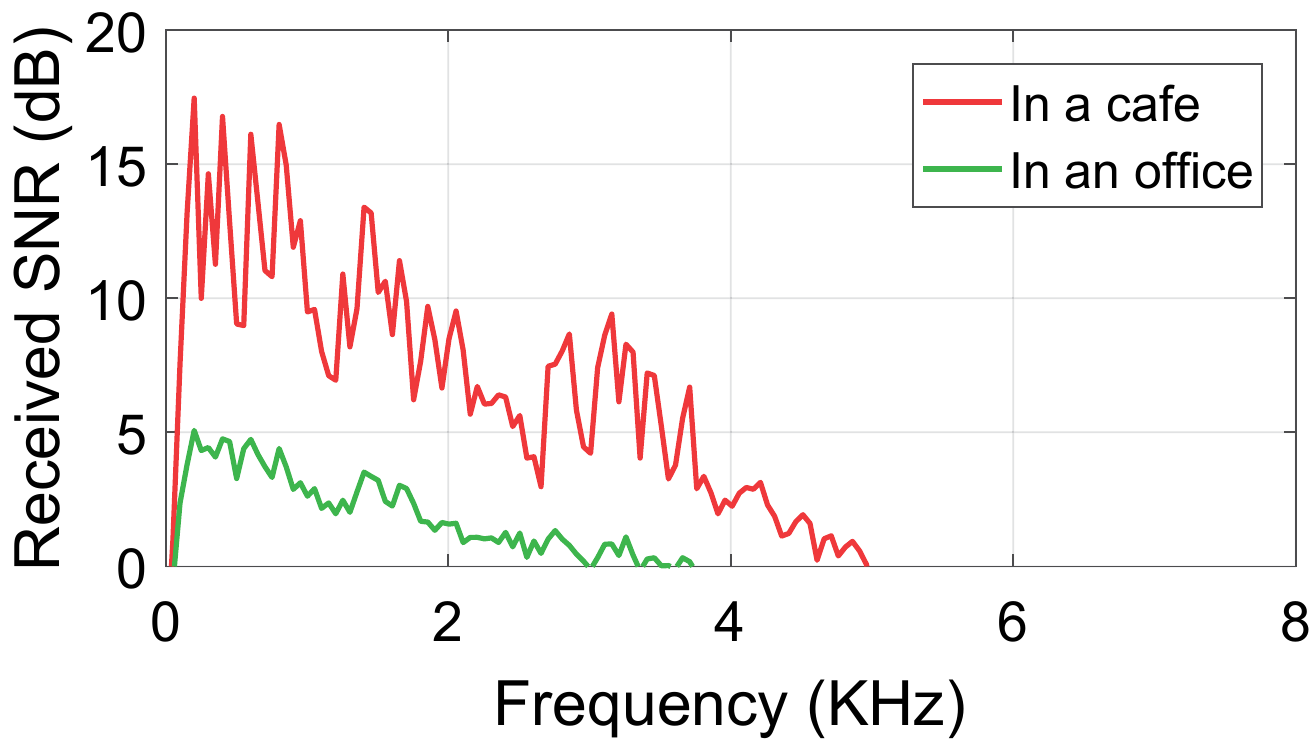}
\vspace{-2mm}
\caption{Spectrum of ambient noise.}
\label{fig:noise}
\end{figure}

\subsubsection{Impact of Smartphone Models and Noise}\label{sec:smartphone}


{
We now evaluate the impact of different smartphone models and ambient noise on the cracking success rate. Figure \ref{fig:noise} shows the energy distribution of ambient noise in a cafe and an office during busy hours, respectively. We can see that most energy of ambient noise resides in low frequency (e.g., less than 5 KHz). } Figure ~\ref{fig:Smartphone-seccess} demonstrates the success rate with 1 sample for two different devices in an office and in a cafe. We can see the success rate of HUAWEI P9 Plus is slightly lower than that of SAMSUNG C9 Pro. This is because P9 Plus possesses a smaller screen (5.5 inches) than that of C9 (6 inches), and the same {line} of the pattern will be shorter on P9 Plus and thus more difficult to be recognized. In addition, we can observe that the ambient noise does not influence the performance of PatternListener obviously. {It is because the generated acoustic signals are in the range of $18 \sim 20$ kHz and ambient noise becomes negligible with these frequencies.} 

\subsubsection{Stealthiness}
{
Finally, we evaluate the stealthiness of PatternListener to analyze the feasibility of the attack in practice. We focus on the rate of CPU consumption, the rate of battery consumption, the size of each audio segment.
We collect the related data from the volunteers' phones when they use their smartphones as usual in 20 days, and then calculate the corresponding values. The main factors related to the stealthiness of malware are shown in Table \ref{tab:stealthiness}. We can see that PatternListener app only consumes extra 4\% CPU cycles during audio signal capturing and the average rate of  battery consumption is only 2.8\%. The reason is that playing/recording the sound incurs low energy cost. PatternListener only monitors the unlocking action and captures the acoustic signals on the phone while pre-processing and pattern reconstruction algorithms do not run on the local phone. In addition, we observe that the average size of each audio sample is only 170.3KB,  which means the network activity is also stealthy.
}

\begin{table}[!t]
\centering
{\scriptsize
\begin{tabular}{c|c|c}
        \Xhline{1.2pt}
            Consumption rate&Consumption rate &Average size of   \\
             of CPU         & of battery      & each sample \\
        \hline
        \hline
           4\%  &2.8\% &170.3KB \\
        \hline

        \Xhline{1.2pt}
\end{tabular}
\caption{The main factors related to stealthiness.}\label{tab:stealthiness}
}
\vspace{-5mm}
\end{table}

%% file: related-work.tex
\section{Related Work}\label{sec:related}

\noindent{\textbf{Pattern Lock Attacks}:}
Smudge attack analyzes the oily residues or smudges left on the screen to infer the unlock pattern~\cite{aviv2010smudge}. However, this approach highly relies on the persistence of the oily residues or smudges which can be easily disturbed by subsequent on-screen activities after unlocking. Zhang et al.~\cite{zhang2016privacy} showed that it is possible to infer the pattern by leveraging the impacts of finger motions on the wireless signals when drawing the pattern. While their approach requires a complex setup and is very easy to be disrupted by moving objects in the environment. Ye et al.~\cite{ye2017cracking} cracked Android pattern lock using video footage that captures the user's fingertip motions as well as part of the device when drawing the pattern. However, the accuracy suffers greatly from filming angle and distance, changes of light, camera shake which are always beyond control. Moreover, it relies on the assumption that the drawing process can be monitored physically, which limits the attack scale of the adversaries. {Aviv et al.~\cite{aviv2012practicality} demonstrated that the accelerometer could be used to learn user gesture-based and tap-based inputs so that they can infer PIN or unlock pattern. However, the proposed approach can only achieve  73\% accuracy of inferring  pattern and 43\% accuracy of inferring PIN with only 50 PINs and 50 patterns.}

\noindent{\textbf{Acoustic Attacks and Tracking}:} Keystroke recognition based on the acoustic emanation has been studied in ~\cite{asonov2004keyboard,berger2006dictionary,zhuang2009keyboard,zhu2014context,wang2014ubiquitous,liu2015snooping}. These approaches leverage the observation that the sound of keystrokes differs slightly from key to key or use time-difference of arrival measurements to identify multiple strokes of the same physical key. In particular, ~\cite{zhu2014context,wang2014ubiquitous,liu2015snooping} employ the advances of mobile devices to identify the keystroke of the nearby keyboard and thus can leverage malicious apps to eavesdrop nearby keyboard input.
Arp et al.~\cite{arpprivacy} explored the capabilities, the current prevalence and technical limitations of embedded ultrasonic beacons in audio and tracked users using the microphone of mobile devices. Trippel et al.~\cite{trippel2017walnut} investigated how analog acoustic injection attacks can damage the digital integrity of the capacitive MEMS accelerometer. To the best of our knowledge, PatternListener is the first work to crack pattern locks using acoustic signals.

Recently, several schemes~\cite{nandakumar2016fingerio,wang2016device,yun2017strata} have been proposed to track 2D gestures by leveraging the smartphone's microphone and speaker. However, they cannot be applied to infer patterns in PatternListener since they require re-configuring smartphone systems to enable 2D finger tracking. These schemes normally utilize the smartphone as an active sonar to identify finger gestures, and the proposed 2D gesture tracking needs to simultaneously use two speaker-microphone pairs. To achieve this, they usually reconfigure the smartphone system to eliminate the impact of the hardware echo cancellation, which is not possible in our attack. Moreover, the accuracy of gesture track is strictly limited by {the region of fingers close to the smartphone.} For example, according to our experiments, we find that the tracking error of LLAP~\cite{wang2016device} is only 0.4 cm when the fingers are in some optimal regions and move 5 cm, while it exceeds 1.6 cm
when the fingers slide on the screen.
In comparison with these schemes, PatternListener can accurately extract the movement features and infer the unlock pattern even if only one speaker-microphone pair is used during pattern drawing on the screen.

\noindent{\textbf{Study of Android Pattern Lock}:}
Uellenbeck et al.~\cite{uellenbeck2013quantifying} studied the security of Android pattern lock and they found that there is a high bias in the pattern selection process. A pilot study on user habits when setting a pattern lock and on their perceptions regarding what constitutes a secure pattern was presented in ~\cite{andriotis2013pilot}. Sun et al.~\cite{sun2014dissecting} analyzed the characteristics of all valid patterns and proposed a way to quantitatively evaluate their strengths. Aviv et al.~\cite{aviv2015bigger} showed that there is a high incidence of repeated patterns and symmetric pairs for both 3 $\times$ 3 and 4 $\times$ 4 patterns. An effective pattern lock strength meter was proposed in ~\cite{song2015effectiveness} to help users choose stronger pattern locks on Android devices. Cho et al.~\cite{chosyspal} proposed a system-guided pattern lock scheme that uses a small number of randomly selected points while selecting a pattern to improve the security of lock patterns.
